\documentclass[trackchanges,twocolumn]{aastex7}

\newcolumntype{C}{>{\centering\arraybackslash}X}
\usepackage{bm}
\usepackage{CJK}
\usepackage{amsmath}
\usepackage{graphicx}
\usepackage{booktabs}
\usepackage{tabularx}

\begin{document}
\begin{CJK*}{UTF8}{gbsn}
\title{Radiation Hydrodynamics of Self-gravitating Protoplanetary Disks:\\I. Direct Formation of Gas Giants via Disk Fragmentation}

\author[orcid=0000-0003-0794-1949]{Yang Ni (倪阳)}
\affiliation{Institute for Advanced Study, Tsinghua University, Beijing 100084, China}
\email[show]{ny22@mails.tsinghua.edu.cn}  

\author[orcid=0000-0001-6858-1006]{Hongping Deng (邓洪平)}
\affiliation{Shanghai Astronomical Observatory, Chinese academy of Science, Nandan Rd 80th, 200030 Shanghai, China}
\email[show]{hpdeng353@shao.ac.cn}  

\author[0000-0001-6906-9549]{Xue-Ning Bai (白雪宁)}
\affiliation{Institute for Advanced Study, Tsinghua University, Beijing 100084, China}
\affiliation{Department of Astronomy, Tsinghua University, Beijing 100084, China}
\email[show]{\\xbai@tsinghua.edu.cn}

\begin{abstract}

Gravitational instability (GI) has long been considered a viable pathway for giant planet formation in protoplanetary disks (PPDs), especially at wide orbital separations or around low-mass stars where core accretion faces significant challenges. However, a primary drawback is that disk fragmentation from GI was generally found to produce over-massive clumps, typically in the mass range of brown dwarfs, although most numerical studies adopted simplified cooling prescriptions or with limited numerical resolution. We conduct a suite of global three-dimensional radiation hydrodynamics (RHD) simulations of self-gravitating PPDs using the meshless finite-mass (MFM) method. By implementing radiation transport via the M1 closure and systematically varying disk mass and opacity, we show that increasing disk mass and lowering opacity promote fragmentation by enhancing radiative cooling. Non-fragmenting disks settle into a gravito-turbulent state with low-order spiral structures and effective angular momentum transport characterized by $\alpha \sim \beta_\mathrm{cool}^{-1}$. In fragmenting disks, a subset of gravitationally bound clumps survives as long-lived fragments. Their initial masses form a consistent distribution around $\Sigma \cdot\lambda_\mathrm{T} \cdot 2 (c_s/\Omega_\mathrm{K})$ (with $\lambda_T$ the Toomre wavelength), corresponding to $\sim 0.3 - 10\,M_\mathrm{J}$ in our simulations, consistent with being gas giants. These results demonstrate that GI can produce planet-mass fragments under more realistic conditions, reinforcing it as a viable gas giant formation pathway and motivating further studies of fragment evolution and observational signatures.

\end{abstract}


\section{Introduction}\label{sec:intro}

The formation of giant planets is a fundamental question in astrophysics, with two primary theoretical pathways proposed: core accretion \citep{Mizuno_1980PThPh..64..544M, Pollack_1996Icar..124...62P, Goldreich_2004ARA&A..42..549G, Lambrechts_2012A&A...544A..32L, Ohno_2021A&A...651L...2O, Pan_2024A&A...682A..89P} and gravitational instability (GI) \citep{Cameron_1978M&P....18....5C, Boss_1997Sci...276.1836B, Boley_2010Icar..207..509B, Forgan_2013MNRAS.432.3168F, Muller_2018ApJ...854..112M, Vigan_2021A&A...651A..72V, Deng_2021NatAs...5..440D}. While core accretion has been the dominant framework, the GI pathway has long been regarded with skepticism. The primary concern is that disk fragmentation was expected to yield over-massive clumps, with birth masses closer to brown dwarfs than to typical gas giants \citep{Forgan2011jeans, Kratter_2016ARA&A..54..271K, Xu_2025ApJ...986...91X}.

These conclusions, however, are largely based on theoretical studies, including analytical estimates and numerical simulations, each with inherent limitations. Analytical estimates cannot capture the highly nonlinear nature of gravitational collapse, and different methods often yield results that differ by up to an order of magnitude \citep[see Section 5.1 of][]{Kratter_2016ARA&A..54..271K}. In terms of numerical simulations, while thermodynamics is known to play a crucial role in regulating fragmentation, as demonstrated in local shearing-box studies \citep{Gammie_2001ApJ...553..174G}, many studies adopted prescribed cooling under various degrees of approximation \citep[e.g.][]{Boley_2006ApJ...651..517B, Zhu_2012ApJ...746..110Z, Baehr_2017ApJ...848...40B, Forgan2017MNRAS.466.3406F, Hirose_2019MNRAS.485..266H, Deng_2021NatAs...5..440D}. Radiation hydrodynamics (RHD) simulations employing flux-limited diffusion (FLD) produced outcomes that are not always consistent with each other, largely because of differing treatments of optically thin regions \citep[e.g.,][]{Boss_2001ApJ...563..367B, Cai_2006ApJ...636L.149C}. More recent simulations using full radiation transport \citep{Xu_2025ApJ...986...91X, Xu_2025ApJ...986...92X} tend to favor the formation of brown-dwarf-mass fragments, but the relatively coarse resolution employed in these calculations may bias the results toward overly massive objects. Crucially, the early, nonlinear stages of fragment collapse must be resolved with sufficient spatial resolution \citep[e.g.][]{Fromang2005effect, Brucy_2021MNRAS.503.4192B}.

On the other hand, gas giants at wide orbital separations \citep{Bohn2020two}, around low-mass stars \citep{Morales_2019Sci...365.1441M, Gan_2022MNRAS.511...83G, Quirrenbach_2022A&A...663A..48Q, lagrange2025evidence}, and in low-metallicity systems \citep{Kanodia_2022AJ....164...81K, Matsukoba_2023MNRAS.526.3933M} challenge standard core accretion theory, as these environments limit the available solid material and growth timescales. These populations suggest that GI may provide a complementary pathway to giant planet formation, at least under specific disk conditions. Moreover, there is growing observational signatures of gravitational instability, including spiral arms in the IM Lup disk \citep{Ueda_2024NatAs...8.1148U, Yoshida_2025NatAs.tmp..193Y}, spiral arms and tentative protoplanets in the AB Aurigae system \citep{Currie_2022NatAs...6..751C, Speedie2024gravitational}, and dusty clumps and fragmenting spirals in the V960 Mon system \citep{Weber2023spirals}.

In this work, we present three-dimensional global radiation hydrodynamics (RHD) simulations of self-gravitating protoplanetary disks, aiming to serve as an initial effort to investigate GI fragmentation with resolved local collapse and first-principle radiation transport. We employ the meshless finite-mass (MFM) method \citep{Hopkins_2015MNRAS.450...53H}, which is a particularly important numerical choice in the context of disk fragmentation. Eulerian methods, based on fixed grids, generally lack the local spatial resolution needed to resolve the runaway collapse of fragments. Adaptive mesh refinement (AMR) can, in principle, provide the necessary resolution, but both AMR and fixed-grid approaches suffer from advection errors: In Cartesian grids, disk rotation is artificially diffused, while in cylindrical grids, it is challenging to follow self-gravitating fragments accurately as they migrate with bulk velocities across the computational domain. Lagrangian approaches, such as smoothed particle hydrodynamics (SPH), naturally track fragment motion and adapt resolution to high-density regions, but they can suffer from limited shock-capturing accuracy and artificial viscosity effects. The MFM method combines the strengths of both: as a Lagrangian Godunov-type scheme, it offers intrinsic adaptivity to collapsing regions while maintaining the accuracy of modern shock-capturing techniques. This makes MFM well-suited for resolving the early, nonlinear stages of fragment collapse and tracking the subsequent evolution of clumps.

In the framework of MFM hydrodynamics, we solve the radiation transport with M1 closure scheme \citep{Levermore_1984JQSRT..31..149L}, with a second-order Harten-Lax-van Leer \citep[HLL,][]{Harten1983} solver, an implicit radiation source term, and an outflow boundary condition that enables thermal radiation to escape naturally from the disk surfaces. Our simulations systematically vary disk mass and opacity to explore how these parameters regulate gravito-turbulence, fragmentation, and the properties of the resulting clumps. Special attention is paid to the initial fragment mass distribution, its connection to local disk conditions, and the survival or disruption of clumps under dynamical interactions. Long-term evolution of the clumps will be presented in a companion paper.

This paper is organized as follows. Section~\ref{sec:method} describes the numerical methods and simulation setups. Section~\ref{sec:results} presents our main findings on disk morphology, gravito-turbulence, clump formation and evolution, and initial fragment masses. In Section~\ref{sec:discussion}, we discuss the implications of our results and compare them with previous studies. Section~\ref{sec:conclusions} summarizes our conclusions.

\section{Method}\label{sec:method}

\subsection{The governing equations}

We perform global hydrodynamic simulations of self-gravitating PPDs using the MFM scheme in the GIZMO code \citep{Hopkins_2015MNRAS.450...53H}. To solve the RHD equations, we have implemented an iterative treatment for stiff source terms and a second-order Riemann solver, including terms up to order $v/c$ \citep[e.g.][]{Skinner_2013ApJS..206...21S, Kannan_2019MNRAS.485..117K, Fuksman_2021ApJ...906...78M}. The implementation details are described in Sections~\ref{sec:rtsolver_transport} and~\ref{sec:rtsolver_source}, and extensive tests are presented in Appendix~\ref{sec:app_a}. The governing equations read

\begin{equation}\label{eq:hydro_rho}
    \frac{\partial \rho}{\partial t} + \nabla \cdot (\rho \bm{v})=0,
\end{equation}

\begin{multline}\label{eq:hydro_p}
    \frac{\partial (\rho\bm{v})}{\partial t} + \rho\bm{v}\cdot \nabla \bm{v}= -\nabla P +\rho \bm{g} + \frac{\rho\kappa_F}{c} \bm{F_r}  \\+(\rho \kappa_{E}E_r\frac{\tilde{c}}{c}-S_r)\frac{\bm{v}}{c}  -\rho \kappa_{E}\frac{\tilde{c}\bm{v}}{c^2}\cdot(E_r\mathbb{I}+\mathbb{P}_r),
\end{multline}

\begin{multline}\label{eq:hydro_e}
    \frac{\partial (\rho e)}{\partial t} + \nabla \cdot [(\rho e + P) \bm{v}]=\rho \bm{v} \cdot \bm{g}  - cS_r + \rho\kappa_E \tilde{c}E_r  \\  +\rho(\kappa_F-2\kappa_E) \frac{\bm{v}}{c}\cdot \bm{F}_r,
\end{multline}

\begin{equation}\label{eq:rt_e}
    \frac{\partial E_r}{\partial t} + \nabla \cdot \bm{F}_r = - \rho\kappa_E \tilde{c}E_r + cS_r  +\rho(2\kappa_E-\kappa_F) \frac{\bm{v}}{c}\cdot \bm{F}_r,
\end{equation}

\begin{multline}\label{eq:rt_f}
    \frac{\partial \bm{F}_r}{\partial t} +\tilde{c}^2 \nabla \cdot \mathbb{P}_r = -\rho \kappa_F  \tilde{c} \bm{F}_r -(\rho \kappa_E E_r\frac{\tilde{c}}{c}-S_r)\tilde{c}\bm{v}  \\ +\rho \kappa_E \frac{\tilde{c}^2}{c} \bm{v}\cdot(E_r\mathbb{I}+\mathbb{P}_r),
\end{multline}

where $\rho$, $\bm{v}$, $P$, and $e=u+\bm{v}\cdot\bm{v}/2$ are the gas density, velocity, pressure, and total specific energy. Here, $u$ is the specific internal energy and is related to the gas pressure via $u=P/[(\gamma -1)\rho]$. In this study, we adopt an adiabatic index of $\gamma = 5/3$, consistent with \citet{Xu_2025ApJ...986...91X}, thereby enabling a more direct comparison with their results. We note, however, that several previous $3$D studies of GI have employed more sophisticated equations of state \citep[e.g.][]{Boss_2007ApJ...661L..73B, Boley_2009ApJ...695L..53B}. Incorporating such treatments would be an important direction for future modeling efforts. The total gravitational acceleration $\bm{g}=\bm{g}_\star+\bm{g}_\mathrm{g}$ includes the gravitational acceleration from the central object, $\bm{g}_\star$, and the gravitational acceleration from the gas itself, $\bm{g}_\mathrm{g}$, which is solved directly via a Tree algorithm \citep{Hopkins_2015MNRAS.450...53H} with adaptive gravitational force softening \citep{price2007energy}.

In Equations (\ref{eq:hydro_p}), (\ref{eq:hydro_e}), (\ref{eq:rt_e}), and (\ref{eq:rt_f}), properties of radiation field are measured in the lab-frame, with all terms up to $v/c$ included. $E_r$, $\bm{F}_r$, and $\mathbb{P}_r$ are respectively the radiation energy density, radiation flux vector, and radiation pressure tensor, defined as angular moments of the specific intensity $I_\nu(\bm{n})$ integrated over solid angle and frequency by
\begin{equation}\label{eq:rt_definition}
    \{\tilde{c}E_r, \bm{F}_r, \tilde{c}\mathbb{P}_r\} = \int_{0}^{+\infty} \int_{4\pi}\{1,\bm{n},(\bm{n} \otimes \bm{n})\} I_\nu(\bm{n}) \mathrm{d}\Omega \mathrm{d}\nu.
\end{equation}

In Equations (\ref{eq:hydro_p}), (\ref{eq:hydro_e}), (\ref{eq:rt_e}), and (\ref{eq:rt_f}),
\begin{equation}\label{eq:sr}
    S_r=\rho\kappa_\mathrm{P} a_r T^4
\end{equation}
is the source term describing the energy exchange between the gas and radiation field, which corresponds to the frequency-integrated gas thermal emission in this study. Here, $a_r=4\sigma_\mathrm{SB}/c$, $\sigma_\mathrm{SB}$ is the Stefan-Boltzmann constant, and $T=[\mu (\gamma-1) u] / k_\mathrm{B}$ is the gas temperature, where $\mu$ is the mean molecular mass, and $k_\mathrm{B}$ is the Boltzmann constant. 

In Equations (\ref{eq:hydro_p}), (\ref{eq:hydro_e}), (\ref{eq:rt_e}), and (\ref{eq:rt_f}), $\kappa_\mathrm{P}$, $\kappa_E$ and $\kappa_F$ are the frequency-integrated specific opacities weighted by the Planck function, energy density, and flux in the comoving frame, respectively. In this study, we do not distinguish between these three opacities, i.e. $\kappa_\mathrm{P}=\kappa_E=\kappa_F=\kappa$, and neglect scattering. We will further discuss the adopted values of $\kappa$ in Section \ref{sec:ppdmodel}.

Equations (\ref{eq:hydro_p}), (\ref{eq:hydro_e}), (\ref{eq:rt_e}), (\ref{eq:rt_f}), and (\ref{eq:rt_definition}) have been reformulated with the reduced speed of light $\tilde{c}$, i.e., adopting the reduced speed of light approximation (RSLA) \citep[e.g.][]{Skinner_2013ApJS..206...21S,Rosdahl_2015MNRAS.449.4380R,Kannan_2019MNRAS.485..117K,Fuksman_2021ApJ...906...78M}. The RSLA enables larger time steps and thus faster simulations, while still preserving the correct steady-state solution of the radiation hydrodynamics equations. However, it introduces caveats: the choice of $\tilde{c}$ must be sufficiently large compared to the characteristic velocities of the problem to ensure accurate propagation of radiation fronts and interactions with matter. The appropriate values of $\tilde{c}$ are problem dependent; in our case, they are tested and further discussed in Section \ref{sec:ppdmodel}.

\subsection{The M1 closure relation}

Equations (\ref{eq:rt_e}) and (\ref{eq:rt_f}) are not complete unless an additional closure relation is introduced. We therefore, adopt the M1 closure relation \cite{Levermore_1984JQSRT..31..149L}, yielding
\begin{equation}
    \mathbb{P}_r = \mathbb{D} E_r,
\end{equation}
where $\mathbb{D}$ is the Eddington Tensor, determined by the local radiation field, 
\begin{equation}
    \mathbb{D} = \frac{1-\chi}{2} \mathbb{I} + \frac{3\chi-1}{2}\bm{n}_F \otimes \bm{n}_F,
\end{equation}
where $\bm{n}_F=\bm{F}_r / |\bm{F}_r|$ is a unit vector in the direction of the flux, and $\chi=(3+4f^2)/(5+2\sqrt{4-3f^2})$ is the Eddington factor. Here,
\begin{equation}
    f=\frac{|\bm{F}_r|}{\tilde{c}E_r}
\end{equation} is the reduced radiation flux. The M1 closure relation reduces to the diffusion limit for $\mathbb{D} \sim \mathbb{I}/3$ when $f \sim 0$, and also captures the free streaming limit for $\mathbb{D} \sim \bm{n}_F \otimes \bm{n}_F$ when $f \sim 1$, with an ad-hoc smooth transition in the intermediate regime.

The RHD framework with M1 closure (M1-RHD) is a widely used approach that has been implemented in various hydrodynamic codes \citep[e.g.][]{Gonzalez_2007A&A...464..429G, Rosdahl_2013MNRAS.436.2188R, Kannan_2019MNRAS.485..117K, Chan_2021MNRAS.505.5784C, Fuksman_2021ApJ...906...78M} and applied across diverse astrophysical contexts \citep[e.g.][]{Rosdahl_2015MNRAS.451...34R, Fuksman_2022ApJ...936...16M, Kannan_2025arXiv250220437K}. Within GIZMO, M1-RHD has also been developed and extensively employed in simulations spanning cosmological, galactic, and local interstellar medium scales \citep[e.g.][]{Lupi_2018MNRAS.474.2884L, Hopkins_2020MNRAS.491.3702H, Grudic_2021MNRAS.506.2199G, Hopkins_2024OJAp....7E..18H}. However, as M1-RHD in the public version of GIZMO was mainly designed for galactic-scale simulations, the transport of radiation is only first-order accurate without any reconstruction at cell interfaces \footnote{In this work, we use the term ``cell'' to denote an individual computational element in the GIZMO simulation. The cell center corresponds to the position of the partition's generating point, while cell interfaces refer to the boundaries shared by two neighboring cells.}, and the source terms are coupled with thermo-chemistry in the multiphase interstellar medium. We therefore adapt the framework of M1-RHD in the public version of GIZMO and independently write our implicit-explicit (IMEX) version of M1-RHD. The major changes include an explicit HLL Riemann solver with piecewise linear spatial reconstruction for radiation transport, an implicit iterative solver for stiff radiation-matter coupling terms, and a geometry-dependent outflow boundary condition for radiation. We elaborate on the implementation of the above three major modifications, respectively, in Section \ref{sec:rtsolver_transport}, \ref{sec:rtsolver_source}, and \ref{sec:rtsolver_boundary}. A series of tests on our M1-RHD module are shown in Appendix \ref{sec:app_a}.

\subsection{Radiation transport via hyperbolic solver}\label{sec:rtsolver_transport}

The transport equation of radiation, i.e., Equations (\ref{eq:rt_e}) and (\ref{eq:rt_f}) with source terms zero out, are hyperbolic conservation laws of photon energy density and photon flux, which can be cast into the conservation form as 

\begin{equation}
    \frac{\partial \mathcal{U}}{\partial t} + \nabla \cdot \mathcal{F}^s = 0,
\end{equation}
where 
\begin{equation}
    \mathcal{U}=\begin{pmatrix} E_r\\ \bm{F}_r \end{pmatrix},
\end{equation}

\begin{equation}
    \mathcal{F}^s(\mathcal{U})=\begin{pmatrix} \bm{F}_r\\ \tilde{c}^2\mathbb{P}_r \end{pmatrix},
\end{equation}
and the superscript ``s'' denotes the variable in the static frame.

Due to the Lagrangian nature of the MFM scheme, each cell interface is not static but moves with the average flow velocity of the neighboring cells  $\bm{w}_\mathrm{LR}=(\bm{v}_\mathrm{L}+\bm{v}_\mathrm{R})/2$. Here, subscripts ``L'' and ``R'' indicate the left and right cell of the interface, respectively. It is convenient to solve for the total flux $\mathcal{F}^m(\mathcal{U})=\begin{pmatrix} \bm{F}_r-E_r\bm{w}_\mathrm{LR}^T\\ \tilde{c}^2\mathbb{P}_r-\bm{F}_r \bm{w}_\mathrm{LR}^T \end{pmatrix}$, which is the sum of the flux over a static interface and the advection due to cell motion \citep{springel2010pur, Hopkins_2015MNRAS.450...53H,Kannan_2019MNRAS.485..117K}. Here, the superscript ``m'' denotes the variable in the moving frame.

Before we calculate the HLL flux over each cell interface, we employ the spatial piecewise linear reconstruction with a similar scheme as in \cite{Kannan_2019MNRAS.485..117K} to achieve higher order accuracy. We first reconstruct $E_r$ and $f$ to the cell interface as
\begin{equation}
    E_{r,\mathrm{face},\mathrm{L}}=E_{r,\mathrm{center},\mathrm{L}} + \nabla E_{r,\mathrm{L}} \cdot \bm{s}_\mathrm{proj,\mathrm{L}},
\end{equation}

\begin{equation}
    f_{\mathrm{face},\mathrm{L}}=f_{\mathrm{center},\mathrm{L}} + \nabla f_{\mathrm{L}} \cdot \bm{s}_\mathrm{proj,\mathrm{L}},
\end{equation}

\begin{equation}
    E_{r,\mathrm{face},\mathrm{R}}=E_{r,\mathrm{center},\mathrm{R}} + \nabla E_{r,\mathrm{R}} \cdot \bm{s}_\mathrm{proj,\mathrm{R}},
\end{equation}

\begin{equation}
    f_{\mathrm{face},\mathrm{R}}=f_{\mathrm{center},\mathrm{R}} + \nabla f_{\mathrm{R}} \cdot \bm{s}_\mathrm{proj,\mathrm{R}},
\end{equation}
where $E_{r,\mathrm{center}}$ is the radiation energy density stored in the cell center, $f_{\mathrm{center}}$ is the reduced radiation flux stored in the cell center, $\nabla E_r$, $\nabla f$ are their gradients, and $\bm{s}_\mathrm{proj}$ is the vector of the distance from the cell center to the face. The subscripts ``L'' and ``R'' refer to the left and right states at each cell interface. To ensure that $f$ remains within $0 \leq f \leq 1$, we employ a minmod slope limiter during the reconstruction of $f$. For $E_{r}$, both the minmod limiter and the more flexible limiter implemented in the public version of GIZMO are available. Unless otherwise stated, we adopt the original limiter from the public release of GIZMO, using the default numerical parameters $\psi_1 = 1/2$ and $\psi_2 = 1/4$ (see Appendix B of \citealt{Hopkins_2015MNRAS.450...53H} for details). We have verified that the choice of slope limiter for reconstructing $E_{r}$ leads to at most moderate differences across all of our tests.
Using the extrapolated quantities, the radiation flux $\bm{F}_r$ at the cell interface is computed as:
\begin{equation}
    \bm{F}_{r,\mathrm{face},\mathrm{L}} = f_{\mathrm{face},\mathrm{L}}\tilde{c}E_{r,\mathrm{face},\mathrm{L}}\bm{n}_{F,\mathrm{L}},
\end{equation}

\begin{equation}
    \bm{F}_{r,\mathrm{face},\mathrm{R}} = f_{\mathrm{face},\mathrm{R}}\tilde{c}E_{r,\mathrm{face},\mathrm{R}}\bm{n}_{F,\mathrm{R}}.
\end{equation}
This reconstruction approach maintains the direction of $\bm{F}_{r}$ from the cell center, while interpolating its magnitude based on the extrapolated $f$ and $E_\mathrm{r}$. As demonstrated by \citet{Kannan_2019MNRAS.485..117K}, this method ensures second-order accuracy for the hyperbolic transport equation, preserves stability, and guarantees that $f_{\mathrm{face}}\in[0,1]$.

The HLL flux over each cell interface is then estimated as
\begin{equation}
    \mathcal{F}^m=
    \left\{
        \begin{array}{ll}
            \mathcal{F}^m_\mathrm{L,face}    & \text{if} \lambda^{-}_\mathrm{m}\geq 0, \\
            \mathcal{F}^m_\mathrm{*}    & \text{if} \lambda^{-}_\mathrm{m}\leq 0 \leq \lambda^{+}_\mathrm{m},\\
            \mathcal{F}^m_\mathrm{R,face}    & \text{if} \lambda^{+}_\mathrm{m}\leq 0, 
        \end{array} 
    \right. 
\end{equation}

where 
\begin{multline}
    \mathcal{F}^m_\mathrm{*}=\frac{\lambda^{+}}{\lambda^{+}-\lambda^{-}}\mathcal{F}^m_\mathrm{L,face}-\frac{\lambda^{-}}{\lambda^{+}-\lambda^{-}}\mathcal{F}^m_\mathrm{R,face}
    \\
    +\frac{\lambda^{+}\lambda^{-}}{\lambda^{+}-\lambda^{-}}(\mathcal{U}_\mathrm{R,face}-\mathcal{U}_\mathrm{L,face}).
\end{multline}
Here, the subscript ``face'' denotes the reconstructed variables, while $\lambda^{+}_\mathrm{m}\equiv \mathrm{max}\{\lambda^{\mathrm{max}}_{\mathrm{L}},\lambda^{\mathrm{max}}_{\mathrm{R}} \}$ and $\lambda^{-}_\mathrm{m}\equiv \mathrm{max}\{\lambda^{\mathrm{min}}_{\mathrm{L}},\lambda^{\mathrm{min}}_{\mathrm{R}} \}$ represent, respectively, the maximum and minimum eigenvalues of the Jacobian $\partial \mathcal{F}^m / \partial \mathrm{U}$ evaluated at the moving cell interfaces. As being proved in \cite{Kannan_2019MNRAS.485..117K}, $\lambda^{+}_\mathrm{m}$ and $\lambda^{-}_\mathrm{m}$ can be converted as 
\begin{equation}
    \lambda^{+}_\mathrm{m} = -\bm{w}_\mathrm{LR}\cdot \bm{\hat{n}}_\mathrm{LR} + \lambda^{+}_\mathrm{s},
    \lambda^{-}_\mathrm{m} = -\bm{w}_\mathrm{LR}\cdot \bm{\hat{n}}_\mathrm{LR} + \lambda^{-}_\mathrm{s},
\end{equation}
where $\bm{\hat{n}}_\mathrm{LR}$ is the face normal of the cell interface, and $\lambda^{\pm}_\mathrm{s}$ are the eigenvalues of the Jacobian $\partial \mathcal{F}^s / \partial \mathrm{U} $ in the static frame, which have been studied in some previous work \citep[e.g.][]{Gonzalez_2007A&A...464..429G, Skinner_2013ApJS..206...21S}{}. Here, we adopt the formulation given in Equation (41a) of \cite{Skinner_2013ApJS..206...21S}. We note that setting $\lambda^{+}_\mathrm{s}=-\lambda^{-}_\mathrm{s}=\tilde{c}$ leads directly to the Global Lax-Friedrichs (GLF) flux, which is also available in our implementation.

We implement our radiation flux transport solver within the original framework of GIZMO, preserving the time integration scheme exactly as used for hydrodynamics and in the public M1-RT module \citep[e.g.,][]{springel2010pur, Hopkins_2015MNRAS.450...53H}. Therefore, we do not repeat those details here.

\subsection{Treatment of stiff radiation source terms}\label{sec:rtsolver_source}

For the radiation source terms, all $\mathcal{O}(v/c)$ contributions are treated explicitly, as their magnitudes remain small in our simulations and do not significantly impact stability. In contrast, potentially stiff source terms—particularly those involving strong absorption—are handled implicitly to ensure numerical stability. In particular, we employ the (semi-)implicit solver on the following equations for radiation source terms:
\begin{equation}\label{eq:hydro_e_source}
    \frac{\partial (\rho u)}{\partial t} = - cS_r + \rho\kappa \tilde{c}E_r,
\end{equation}

\begin{equation}\label{eq:rt_e_source}
    \frac{\partial E_r}{\partial t}  = - \rho\kappa \tilde{c}E_r + cS_r,
\end{equation}

\begin{equation}\label{eq:rt_f_source}
    \frac{\partial \bm{F}_r}{\partial t} = -\rho \kappa  \tilde{c} \bm{F}_r.
\end{equation}

Following \cite{Skinner_2013ApJS..206...21S}, we solve Equation (\ref{eq:rt_f_source}) using a standard $\theta$-scheme and update $\bm{F}_r$ as
\begin{equation}
    \bm{F}_r^{n+1}=\bm{F}_r^{n}[\frac{1-(1-\theta)\tilde{c}\rho\kappa \Delta t}{1+\theta \tilde{c} \rho \kappa \Delta t}],
\end{equation}
where $\bm{F}_r^{n+1}$ is the radiation flux vector to be solved on the $(n+1)$th step, $\bm{F}_r^{n}$ is the radiation flux vector on the $n$th step, $\theta=1$ translates into the Backwards Euler Method, and $\theta=1/2$ translates into the Trapezoidal Method. By default, we set $\theta=0.51$ unless otherwise specified.

Then, solving Equations (\ref{eq:hydro_e_source}), (\ref{eq:rt_e_source}), and Equation (\ref{eq:sr}) simultaneously with a $\theta$-scheme, we have
\begin{equation}\label{eq:stiffsourceroot}
    c_4 (U^{n+1})^4 + c_1 U^{n+1} + c_0 =0,
\end{equation}
where $U^{n+1}=\rho u^{n+1}$ is the internal energy to be solved at the $n+1$ step. The coefficients are defined as $c_4=\rho \kappa c \Delta t a_r [\frac{\mu(\gamma-1)}{\rho k_B}]^4\theta$, $c_1=\rho \kappa \tilde{c} \Delta t \theta +1$, $c_0=(1-\theta)\rho\kappa \Delta t \{c a_r  [\frac{\mu(\gamma-1)}{\rho k_B}]^4 - \tilde{c}E_r^n \}-(U^n+\rho \kappa \Delta t \tilde{c} \theta 
\bar{E})$, where $\Delta t$ is the current timestep, $U^n$ is the internal energy at step $n$, $E_r^n$ is the radiation energy density at step $n$, $\bar{E}=U^n+E^n=U^{n+1}+E^{n+1}$ is the conserved energy during matter-radiation coupling, and $E^{n+1}$ is the radiation energy density to be solved at the $n+1$ step. We then adopt the same root-finding strategy to solve Equation (\ref{eq:stiffsourceroot}) for $U^{n+1}$ as in \cite{Skinner_2013ApJS..206...21S}, and subsequently obtain $E^{n+1}$ from energy conservation.

\subsection{Radiation outflow boundary conditions}\label{sec:rtsolver_boundary}

To model thermal radiation escaping from the disk surface to infinity, which effectively imposes an outflow boundary condition for radiation, Eulerian simulations typically adopt ghost cells with either a constant low radiation energy density and/or a zero-gradient radiation flux \citep[e.g.][]{Flock_2013A&A...560A..43F,Fuksman_2022ApJ...936...16M,Zhang_2024ApJ...968...29Z}{}. However, due to the Lagrangian nature of the MFM scheme and the highly dynamic surface of gravitationally unstable disks, implementing ghost cells at the disk surface to allow photons to escape without interfering with fluid dynamics is not straightforward. Previous Lagrangian simulations, such as those using SPH, often relied on the geometric arrangement of particles within their smoothing kernels to identify surface particles, onto which additional cooling terms were applied \citep{Mayer_2007ApJ...661L..77M}. A drawback of such an approach is the introduction of at least one free parameter for identifying surface particles.

Alternatively, taking advantage of MFM's well-defined cell interfaces—an improvement over traditional SPH—we compute the solid angle coverage for each cell using the face area vectors of all interfaces within its kernel. This approach has been widely adopted in astrophysical simulations using GIZMO \citep[e.g.,][]{Hopkins_2018MNRAS.477.1578H, Hopkins_2019MNRAS.483.4187H} and other quasi-Lagrangian codes such as \textsc{AREPO} \citep[e.g.,][]{Smith_2018MNRAS.478..302S, Marinacci_2019MNRAS.489.4233M}. For a given cell $P_i$ and a neighboring cell $P_j$ within its kernel, the interface between them is denoted by $A_{i,j}$. We approximate an oblique cone whose base is $A_{i,j}$ and apex is located at $P_i$, and compute the solid angle $\Omega_{i,j}$ subtended by this interface. The geometric formulation and exact computation of $\Omega_{i,j}$ in GIZMO are described in detail by \citet{Hopkins_2018MNRAS.477.1578H} (see also Appendix E of \citet{Hopkins_2018MNRAS.480..800H}), to which we refer readers for further information. Summing over all such neighbors yields the total solid angle $\Omega_{i,\mathrm{total}} = \sum_{j=1}^{N_\mathrm{kernel}} \Omega_{i,j}$. For cells located well within the disk, $\Omega_{i,\mathrm{total}}$ should approach $4\pi$. We confirm this in the fiducial run of our simulations: for cells near the disk midplane (defined by $|z|/\sqrt{x^2 + y^2} < 0.05$), we find an average solid angle of $\left\langle \Omega_{\mathrm{total}}/(4\pi) \right\rangle = 4.00 \pm 0.11$, which varies only slightly throughout the simulation. In contrast, for surface cells with sparse neighbors in the direction perpendicular to the disk surface, $\Omega_{i,\mathrm{total}}$ is significantly less than $4\pi$. We define the missing solid angle as $\Delta \Omega_\mathrm{empty}=(4\pi-\Omega_{i,\mathrm{total}})$, which estimates the open solid angle available for radiation to escape from $P_i$ to infinity. To account for radiation losses through this open angle, we first update the boundary cells using the standard flux exchanges with all their neighboring cells. After this update, we introduce additional source terms that further modify the radiation energy density and flux as follows:
\begin{multline}\label{eq:corr_er}
    \int E_r^{n+1}\mathrm{d}V_\mathrm{cell} = 
    \int E_r^{n} \mathrm{d}V_\mathrm{cell} \\ 
    - \tilde{c} E_r^{n} \cdot e^{-\rho \kappa L_\mathrm{cell}} \Delta \Omega_\mathrm{empty} L_\mathrm{cell}^2 \Delta t,
\end{multline}
\begin{multline}\label{eq:corr_fr}
    \int \bm{F}_r^{n+1}\mathrm{d}V_\mathrm{cell} = \int\bm{F}_r^{n} \mathrm{d} V_\mathrm{cell} \\
    + \tilde{c}^2 E_r^{n} \cdot e^{-\rho \kappa L_\mathrm{cell}} \Delta \Omega_\mathrm{empty} L_\mathrm{cell}^2 \Delta t \bm{\hat{n}}_\mathrm{out} ,
\end{multline}
where $L_\mathrm{cell}$ is the physical size of the cell, $V_\mathrm{cell}$ is the physical volume of the cell, $\Delta t$ is the current timestep, $E_r^{n+1}$ is the radiation energy density to be solved on the $n+1$ step, $E_r^{n+1}$ is the energy density on the $n$ step, $\bm{F}_r^{n+1}$ is the radiation flux vector to be solved on the $n+1$ step, $\bm{F}_r^{n}$ is the radiation flux vector on the $n$ step, $\bm{\hat{n}}_\mathrm{out}$ is the direction of sparest cell distribution within one kernel. Although cells may have irregular shapes, we approximate their characteristic size as
$L_\mathrm{cell}=[4 \pi h^3/(3 N_\mathrm{ngb})]^{1/3}$, where $h$ is the smoothing length of the cell's kernel and $N_\mathrm{ngb}=32$ is the number of neighboring cells contained within one kernel. These corrections assume that radiation escapes through the cone subtended by $\Delta \Omega_\mathrm{empty}$ in a free-streaming manner with $f\sim 1$. It is important to note that this radiation boundary treatment is a compromise. Lagrangian codes cannot resolve arbitrarily high, optically thin regions of the disk, and our cone-based estimation assumes circular bases for ${A}_{i,j}$, whereas in reality, these interfaces are polygons. Nonetheless, Equations (\ref{eq:corr_er}) and (\ref{eq:corr_fr}) only significantly affect regions where $\rho \kappa L_\mathrm{cell} \ll 1$ and $\Omega_\mathrm{empty}/\pi \sim 1$, which translates into less than one percent of cells in all of our simulations. In these regions, cells are extremely sparse, and the fluid dynamics are poorly resolved. These layers are not the focus of our study. Instead, our primary interest lies in the well-resolved dense regions, such as the disk midplane and gravitationally collapsing clumps. Moreover, the convergence we observe in our quadra-resolution simulations suggests that this approximate boundary treatment does not compromise the accuracy of our main results.

\subsection{Disk setup and initial conditions} \label{sec:ppdmodel}

In this subsection, we describe the setup of the PPD model that serves as the basis for our simulations. Earlier studies found that starting from a non-turbulent disk may lead to spurious fragmentation \citep{Paardekooper_2011MNRAS.416L..65P,Deng_2017ApJ...847...43D}, so we adopt gravito-turbulent disks as the initial conditions to avoid this artifact. We first use rejection sampling to construct a three-dimensional disk model assuming vertical hydrostatic equilibrium, temporarily ignoring disk self-gravity. We then run the disk model with self-gravity for $5$ outer orbit periods, bringing it to a gravito-turbulent state by cooling at characteristic time scales of $\beta_{\mathrm{cool}}\Omega_\mathrm{K}^{-1}$~\citep{Gammie_2001ApJ...553..174G}, where $\Omega_\mathrm{K}$ is the Keplerian angular frequency. Below, we detail the initial smooth disk and constant $\beta$ cooling process.

 We adopt global disk models with power-law surface density and temperature profiles similar to previous studies \citep{Lodato_2004MNRAS.351..630L,Meru_2011MNRAS.411L...1M,Deng_2020ApJ...891..154D}. The disk expands from $30\,\mathrm{AU}$ to $100\,\mathrm{AU}$, following surface density and temperature profiles as $\Sigma\propto R^{-1}$ and $T \propto R^{-1/2}$, respectively. The mass of the central star is fixed at $M_{\star}=1M_\odot$ in all simulations. The disk masses $M_\mathrm{d}$ in all simulations are below $0.2M_\odot$ (Table \ref{tab:setup}), each resolved with $2.1$ million equal-mass cells, corresponding to a minimum resolved mass of $\lesssim 10^{-4},M_\mathrm{J}$. The initial temperature is chosen so that the Toomre $Q=c_s \Omega_\mathrm{K} /(\pi G \Sigma)$ parameter is $1.5$ at $100\,\mathrm{AU}$, where $c_s$ is the gas sound speed. The disk is overall stable, while the outer disk is on the verge of gravitational instability, saving simulation time in the transition phase. We then evolve the axisymmetric smooth disks with cooling, assuming  $\beta_\mathrm{cool}=10$,  for $5\,\mathrm{kyr}$ and eventually generate saturated gravito-turbulent disk models, as the initial conditions for our RHD simulations.

In all disk simulations, we assume a simple opacity \citep{Xu_2023ApJ...946...94X}
\begin{equation} \label{eq:kappa}
\kappa=\left\{
\begin{array}{rcl}
  (\frac{T}{100\mathrm{K}})^2 \kappa_0  &(T<100\mathrm{K}) \\
 \kappa_0  &(T\geq 100\mathrm{K})
\end{array} \right. 
\end{equation}
We explored two opacities characterized by $\kappa_0=1\mathrm{cm}^2\cdot \mathrm{g}^{-1}$ and $\kappa_0=5\mathrm{cm}^2\cdot \mathrm{g}^{-1}$ (see Table \ref{tab:setup}).
We note that at higher temperatures, more complex models would predict several sharp transitions in the opacity due to the sublimation of different dust components \citep[e.g.][]{Bell_1994ApJ...427..987B, Pollack_1994ApJ...421..615P, Semenov_2003A&A, Malygin_2014A&A...568A..91M}. However, given that our simulations primarily probe the cold outer disk regions ($T \sim 10-20 \,\mathrm{K}$), where dust opacities vary slowly with temperature, Equation (\ref{eq:kappa}) provides an adequate approximation.

The reduced speed of light is fixed at $\tilde{c}/c=4\times 10^{-4}$, satisfying the RSLA static diffusion criterion \citep[e.g.][]{Skinner_2013ApJS..206...21S, Rosdahl_2015MNRAS.449.4380R}{}
\begin{equation}\label{eq:rsla}
    \tilde{c} >10 |v_\mathrm{z}| \max \{1,\tau_\mathrm{mid} \}.
\end{equation}
We quadrupled the reduced speed of light and found little change compared to the fiducial simulation (see Appendix~\ref{sec:app_b}). Finally, we summarize all simulation parameters in Table \ref{tab:setup}.

\begin{deluxetable}{lccc}
\tablecaption{Summary of simulation setups\label{tab:setup}}
\tablehead{
\colhead{Name} & 
\colhead{Disk mass $M_\mathrm{d}$ ($M_{\odot}$)} &
\colhead{Opacity $\kappa_0$ (cm$^2$ g$^{-1}$)} &
\colhead{Simulation duration $T$ (kyr)}
}
\startdata
m13k1 & 0.13 & 1 & 5.0 \\
m15k1 & 0.15 & 1 & 3.4 \\
m17k1 & 0.17$^{\dagger}$ & 1 & 1.6 \\
m20k1 & 0.20 & 1 & 1.2 \\
m13k5 & 0.13 & 5$^{\dagger}$ & 5.0 \\
m15k5 & 0.15 & 5$^{\dagger}$ & 5.0 \\
m17k5 & 0.17$^{\dagger}$ & 5$^{\dagger}$ & 1.8 \\
m20k5 & 0.20 & 5$^{\dagger}$ & 1.7 \\
\enddata
\tablecomments{The disk mass $M_\mathrm{d}$ is measured at the initial gravito-turbulent state. Simulations are run for $5\,\mathrm{kyr}$ at maximum. For simulations that experience clump formation leading to prohibitively small timesteps, they are stopped once the minimum timestep drops below $ 10^{-5}\,\mathrm{yr}$. The superscript ${\dagger}$ denotes the fiducial values of $M_\mathrm{d}$ and $\kappa_0$.}
\end{deluxetable}

\subsection{Disk diagnostics}

To characterize the physical properties of our disks, we divide the disk into $30$ linearly spaced radial bins between $30$ AU and $100$ AU. The surface density $\Sigma$ is derived by summing up the mass of all particles within a certain annulus and dividing it by the area of the annulus. We then compute the mass-averaged temperature $T=\langle T \rangle_m$ and the midplane optical depth $\tau_\mathrm{mid}=\langle \kappa \rangle_m \cdot \Sigma/2$, where the averages are taken over all gas cells in the bin. Here, $\langle \kappa \rangle_m$ denotes the mass-weighted mean opacity.

It is also useful to compare our RHD simulations with previous simulations with the simplified $\beta_{\mathrm{cool}}$ cooling prescription, where the turbulence property \citep{Cossins_2009MNRAS.393.1157C,Deng_2020ApJ...891..154D,Brucy2021two,Bethune_2021A&A...650A..49B} and disk fragmentation \citep{Meru_2011MNRAS.411L...1M,Deng_2017ApJ...847...43D} have been extensively studied. To that end, an dimensionless cooling efficiency $\beta_\mathrm{cool}$ in the RHD simulations is defined as
\begin{equation}\label{sec:betacool}
    \beta_\mathrm{cool}=\Omega_\mathrm{K} \frac{\int \rho u\mathrm{d}V}{\int_s \bm{F}_{r} \cdot \mathrm{d}S}
    =\Omega_\mathrm{K} \frac{\sum_{j=1}^{N_{\mathrm{ann}}} m_{\mathrm{cell},j} u_j}{\sum_{j=1}^{N_{\mathrm{ann}}} V_{\mathrm{cell},j}\nabla \cdot \bm{F}_{r,j}},
\end{equation}
where the Keplerian angular velocity $\Omega_\mathrm{K}$ is calculated at the center of each radial bin, the summation is performed over all gas cells in each radial bin, $N_{\mathrm{ann}}$ is the number of cells in each radial bin, and $m_{\mathrm{cell},j}$ is the mass of the $j$-th cell (cells have equal masses here).

To quantify the efficiency of angular moment transport within disks, we compute the radial profiles of the volume-averaged thermal pressure $\langle P \rangle _V$, volume-averaged Reynolds stress $\langle T_{r\phi}^{\mathrm{Rey}} \rangle_V= \langle \rho \delta v_r \delta v_\phi \rangle_V$, and volume-averaged gravitational stress $ \langle T_{r\phi}^{\mathrm{Grav}} \rangle_V= \langle \delta g_r \delta g_\phi/(4 \pi G) \rangle_V$, where $\delta v_r$, $\delta v_\phi$, $g_r$, $g_\phi$ are the fluctuations of the radial/azimuthal velocity/gravitational acceleration \citep{Lynden-Bell_1972MNRAS.157....1L}, respectively. The effective viscosity parameter \citep{Shakura_1973A&A....24..337S, Lynden-Bell_1974MNRAS.168..603L} reads,
\begin{equation}
    \alpha=(\langle T_{r\phi}^{\mathrm{Rey}} \rangle_V+\langle T_{r\phi}^{\mathrm{Grav}} \rangle_V)/\langle P \rangle _V,
\end{equation}
where the averaging is performed over all gas cells in each radial bin.

To analyze non-axisymmetric structures in the disk, we compute the azimuthal power spectrum of the gas density. Within each radial bin, we perform a discrete Fourier transform of the volumetric density in the azimuthal direction, following the approach of \citet{Cossins_2009MNRAS.393.1157C} and \citet{Deng_2020ApJ...891..154D}. For a given radial bin, the normalized amplitude of the $m$-th density mode is calculated as
\begin{align}
A_m  &=\frac{1}{ \sum_{j=1}^{N_{\mathrm{ann}}} m_{\mathrm{cell},j}}\left| \sum_{j=1}^{N_{\mathrm{ann}}} m_{\mathrm{cell},j}  e^{-im\phi_j} \right|  \notag\\ 
 &= \frac{1}{N_{\mathrm{ann}} }\left| \sum_{j=1}^{N_{\mathrm{ann}}}   e^{-im\phi_j} \right|,
\end{align}
where the summation is performed over all gas cells in each radial bin, and $\phi_j$ is the azimuthal angle of the $j$-th cell. This allows us to quantify the strength and radial dependence of spiral features.

\subsection{Clump tracking}

In the post-processing analysis, we identify clumps formed in the gravitationally unstable disk at each epoch using \texttt{CloudPhinder}\footnote[1]{\url{https://github.com/mikegrudic/CloudPhinder}}, a structure-finding algorithm which starts from local density peaks and identifies the largest gravitational binding structures by searching for neighboring cells. At any time $t$, a \emph{clump} is defined as a marginally bound structure with a virial parameter
\begin{equation}\label{eq:defalpha}
  \alpha_\mathrm{vir}(t) = \frac{2[E_\mathrm{K,clump}(t) + U_\mathrm{clump}(t)]}{|E_\mathrm{P,clump}(t)|} < \alpha_\mathrm{crit} = 2,
\end{equation}
where $E_\mathrm{K,clump}(t)$, $U_\mathrm{clump}(t)$, and $E_\mathrm{P,clump}(t)$ are the kinetic, internal, and gravitational energies of the structure at the $t$ moment, respectively. A structure is gravitationally bound if $\alpha_\mathrm{vir} \leq 1$. The threshold $\alpha_\mathrm{crit}$ is a free parameter that determines whether a structure is considered to be an over-density of interest. Here, we adopt a constant $\alpha_\mathrm{crit} = 2$ to include marginally bound clumps so that we can further track their transition toward disruption or fully bound states. We further discuss the exact evolution of $\alpha_\mathrm{vir}(t)$ in Section~\ref{sec:clumpevolution}. We compute the effective radius of a clump as
\begin{equation}
R_{\mathrm{eff}}(t) = \sqrt{\frac{5}{3} \cdot \frac{\sum_i m_{\mathrm{cell},i} [\bm{r}_{i}(t) - \bm{r}_c(t)]^2}{\sum_i m_{\mathrm{cell},i}}}, \quad i \in \text{clump},
\end{equation}
where $\bm{r}_i(t)$ is the position of the $i$-th cell within the clump at the $t$ moment, and $\bm{r}_c(t)$ is the center of mass of the clump at the time $t$. The factor $5/3$ originates from the moment of inertia of a uniform-density sphere. In our context, this serves as an approximation for the geometry of dense clumps rather than an exact relation. We also compute the clump mass at the $t$ moment,
\begin{equation}
m_\mathrm{clump}(t) = \sum_i m_{\mathrm{cell},i}, \quad i \in \text{clump},
\end{equation}
where $\bm{r}_c(t)$ is the center of mass of the clump at the $t$ moment.

Once clumps are identified in each simulation snapshot, we track their evolution over time, taking advantage of the unique IDs of their constituent cells/particles. Specifically, a clump in the current snapshot is associated with a clump in the previous snapshot if they share the majority of their cell IDs, indicating continuity in mass content \citep[see][for more details]{Ni_2025A&A...699A.282N}. Thanks to the high cadence of our simulation outputs ($t_\mathrm{output} = 10\,\mathrm{yr}$), the overlap in cell IDs for clumps in consecutive snapshots typically exceeds $50\%$. As a result, the exact threshold for determining matches has little impact on the robustness of the tracking, and visual inspection further confirms that the clump tracking is robust regardless of minor details in the algorithm's configuration.

Finally, we define \emph{fragments} as a subset of clumps. A clump is characterized as a \emph{fragment} only if it (1) survives without disruption until the end of the simulation and (2) reaches $\min[\alpha_\mathrm{vir}(t)] < 1.1$ during its evolution. These objects are considered long-lived, gravitationally collapsing structures. We define the initial fragment mass $m_\mathrm{frag}$ as
\begin{equation}
    m_\mathrm{frag} = m_\mathrm{clump}(t=t_\mathrm{identify}),\text{clump}\in\text{fragments},
\end{equation}
where $t_\mathrm{identify}$ is the time $t$ that this clump is identified via the criteria of Equation (\ref{eq:defalpha}). Here, $m_\mathrm{frag}$ is not a function of time, but identical for each fragment to capture the mass during its initial collapse process.

\section{Results}\label{sec:results}

\subsection{Overview}\label{sec:overview}

\begin{figure*}[ht!]
\plotone{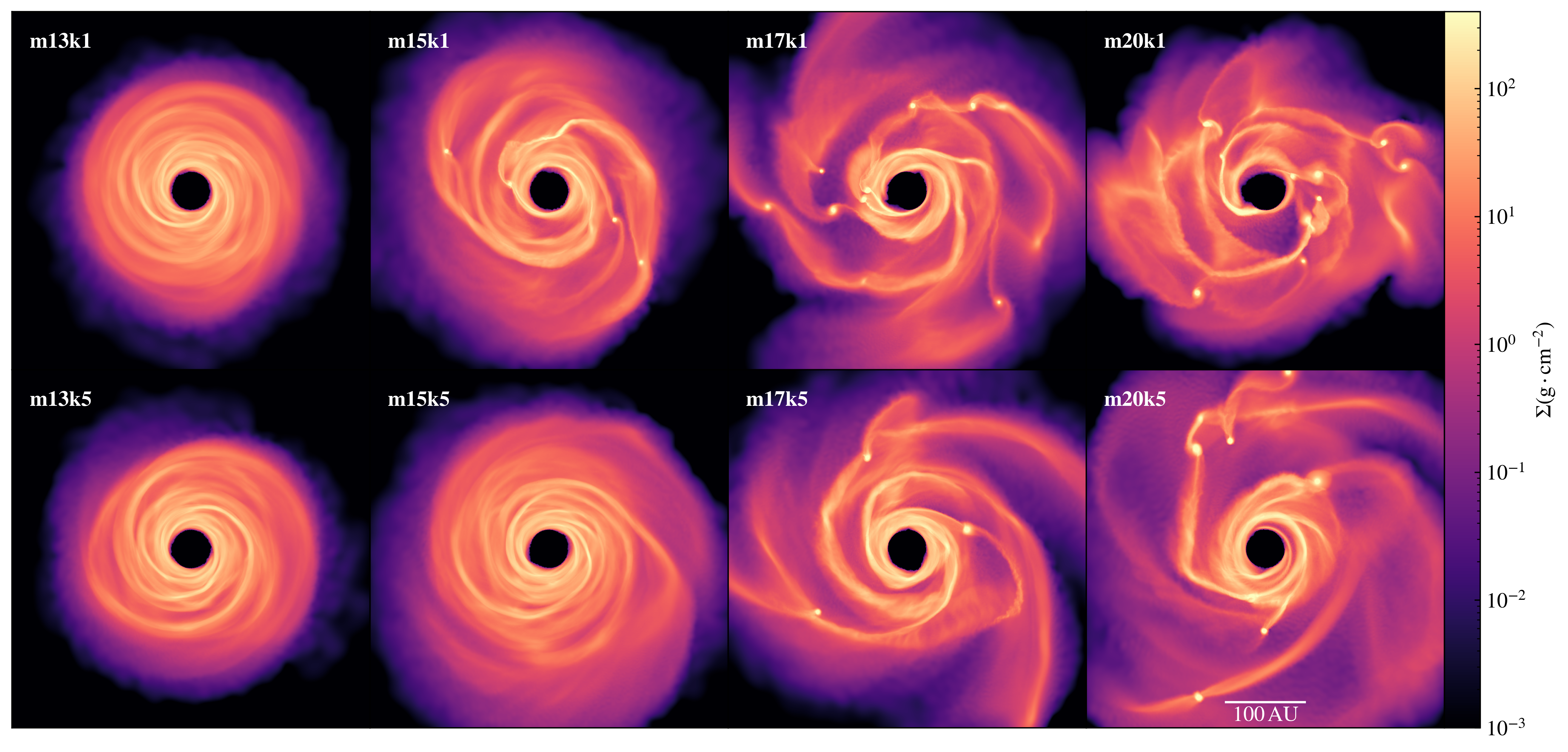}
\caption{The gas surface density maps at the end of the simulations as labelled in the upper left. The box size is $225\,\mathrm{AU}\times225\,\mathrm{AU}$, and a scale bar of 100 AU is included. 
\label{fig:RhoProj_multirun}}
\end{figure*}

Figure~\ref{fig:RhoProj_multirun} presents the gas surface density map at the end of the simulations (see Table \ref{tab:setup}) in a $225\,\mathrm{AU} \times 225\,\mathrm{AU}$ domain. In the fiducial-opacity runs (the bottom four panels), the lower-mass disk \texttt{m13k5} shows relatively symmetric spiral arms in the gravito-turbulence, with no signs of fragmentation. As the disk mass increases, the \texttt{m15k5} run develops more prominent spirals with larger pitch angles; It also experiences significant viscous spreading in the outer disk, extending to larger radii. In the high-mass regime (\texttt{m17k5} and \texttt{m20k5}), however, the disks evolve into markedly asymmetric configurations, where distinct dense, compact regions emerge due to gravitational collapse and fragmentation. When a substantial number of clumps emerge in the fragmented disks, the gravitational interactions between the disk and clumps lead to a complex dynamical interplay. In these cases, the disk structure is no longer dominated by coherent spiral arms; instead, clump-disk interactions and clump-clump close encounters play a key role in shaping the overall morphology\citep[see also][]{kubli2025stochastic}. In the corresponding low-opacity runs (the upper four panels), we observe broadly similar fragmentation trends, though the threshold for collapse shifts to slightly lower disk masses: The \texttt{m13k1} run maintains quasi-steady gravito-turbulent states with prominent spiral arms but shows no evidence of collapse. As the disk mass increases, fragmentation emerges: the \texttt{m15k1}, \texttt{m17k1}, and \texttt{m20k1} runs all show the development of dense clumps and disrupted disk morphology.

\begin{figure*}[ht!]
\plotone{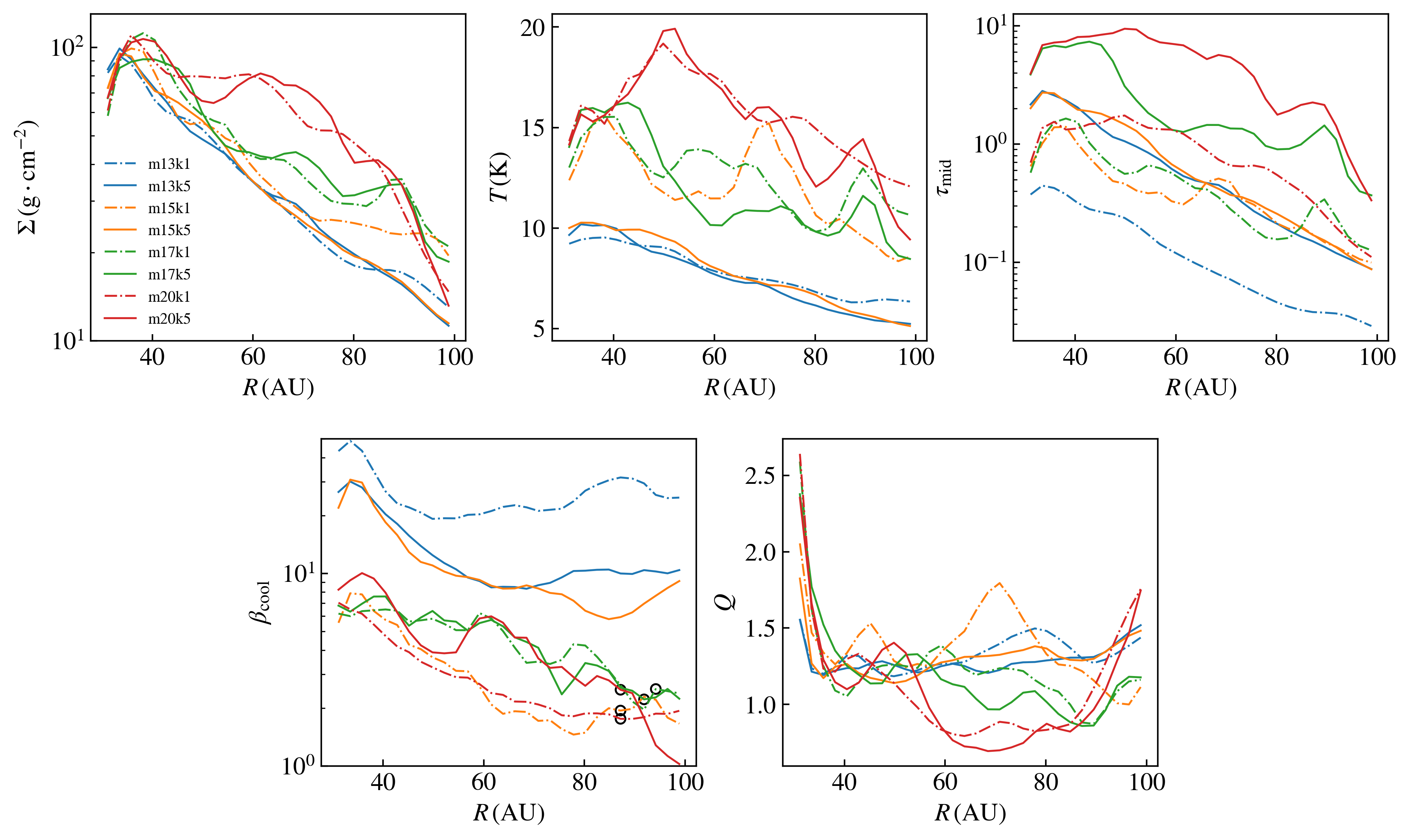}
\caption{Time-averaged radial profiles of $\Sigma$, $T$, $\tau_\mathrm{mid}$, $\beta_\mathrm{cool}$, $Q$ in all the simulations. Profiles of the three non-fragmented disks, i.e., m13k1, m13k5, and m15k5, are time-averaged during the last $1\,\mathrm{kyr}$ of the simulation ($4-5\,\mathrm{kyr}$) to capture the GI-saturated quasi-steady state. Profiles of fragmented disks, i.e., m15k1, m17k1, m17k5, m20k1, and m20k5, are time-averaged during the last $0.1\,\mathrm{kyr}$ before the first clumps in these simulations emerge to capture the instantaneous disk properties generating the fragmentation. Black circles indicate locations of the first clumps in the fragmented disks.
\label{fig:profiles_all}}
\end{figure*}

To gain quantitative insight into the physical conditions that distinguish gravito-turbulent from fragmenting disks, we examine the time-averaged radial profiles of surface density $\Sigma$, temperature $T$, midplane optical depth $\tau_\mathrm{mid}$, dimensionless cooling efficiency $\beta_\mathrm{cool}$, and Toomre $Q$, shown in Figure~\ref{fig:profiles_all}. We note that these profiles are not measured at identical simulation times. For the non-fragmenting disks (\texttt{m13k1}, \texttt{m13k5}, \texttt{m15k5}), profiles are averaged over the final $1\,\mathrm{kyr}$ of evolution ($4 - 5\,\mathrm{kyr}$), reflecting the quasi-steady gravito-turbulent state. For the fragmenting cases (\texttt{m15k1}, \texttt{m17k1}, \texttt{m17k5}, \texttt{m20k1}, \texttt{m20k5}), profiles are averaged over the last $0.1\,\mathrm{kyr}$ before the emergence of the first clumps, capturing the disk conditions at the onset of fragmentation.

In the gas surface density profiles (top-left panel), non-fragmenting disks (\texttt{m13k1}, \texttt{m13k5}, \texttt{m15k5}) exhibit smooth, gradually decreasing profiles, consistent with a self-regulated gravito-turbulent state. In such a state, the Toomre $Q$ parameter remains close to (slightly above) unity across most of the disk. This condition implies $c_s \Omega_\mathrm{K}/\Sigma \sim \mathrm{constant}$, so that, with $\Omega_\mathrm{K} \propto R^{-3/2}$, one expects $c_s/\Sigma \propto R^{3/2}$, or equivalently $T^{1/2}/\Sigma \propto R^{3/2}$. In Figure~\ref{fig:profiles_all}, the measured profiles indeed follow this scaling: $\Sigma \propto R^{-2}$ and $T \propto R^{-1}$ beyond $R \gtrsim 40\,\mathrm{AU}$. The temperature profile, however, is slightly shallower than $R^{-1}$, which causes $Q$ to rise mildly at large radii.

In contrast, fragmenting disks (\texttt{m15k1}, \texttt{m17k1}, \texttt{m17k5}, \texttt{m20k1}, \texttt{m20k5}) display enhanced fluctuations, especially in the outer regions ($R \gtrsim 60$ AU), where local mass accumulation in spiral arms or clumps produces noticeable deviations from the smooth profile. At fixed opacity, increasing the disk mass leads to systematically higher surface densities, as expected. At fixed mass, comparisons between \texttt{m13k1} and \texttt{m13k5}, \texttt{m17k1} and \texttt{m17k5}, and \texttt{m20k1} and \texttt{m20k5} show similar surface density distributions within each pair. A notable exception is \texttt{m15k1}, whose fragmented state exhibits a significantly denser outer disk compared to the non-fragmented \texttt{m15k5} disk, suggesting an enhanced buildup of mass before collapse.

The temperature profiles (top-middle panel) show that non-fragmenting disks remain cool throughout, particularly in their outer regions where $T \lesssim 10\,\mathrm{K}$. This reflects a balance between gravitational heating and radiative cooling in the saturated gravito-turbulent regime. By contrast, fragmenting disks develop significantly higher temperatures, with local peaks up to $\sim 20\,\mathrm{K}$. This enhanced heating arises from the rapid conversion of gravitational potential energy into thermal energy during local collapse and from strong, localized shocks associated with clump-disk interactions, in contrast to the more distributed, weaker shocks in gravito-turbulent spirals. For example, despite having the same total mass, the fragmented \texttt{m15k1} run is systematically hotter than the non-fragmented \texttt{m15k5} run by a factor of $\sim 1.5$ in the outer disk. The \texttt{m15k1} disk also maintains a slightly higher surface density, reflecting mass pile-up at the onset of fragmentation. Taken together—slightly larger $\Sigma$, $\sim 1.5$ times higher $T$, and a roughly comparable $\tau_\mathrm{mid}$ (see next paragraph)—the \texttt{m15k1} disk reaches cooling times shorter by about a factor of five relative to \texttt{m15k5}, pushing it into the fragmenting regime.

As shown in the midplane optical depth (top-right panel) profiles, our model variations cover the optical depth ranging from a mostly optical thick disk (\texttt{m20k5}) to an optically thin disk (\texttt{m13k1}). In our simulations, the opacity follows a temperature-dependent prescription $\kappa \propto T^2$ for $T < 100,\mathrm{K}$, so $\tau_\mathrm{mid} \propto \kappa_0 \Sigma\,T^2$. For the following three of the four same-mass pairs: \texttt{m13k1} and \texttt{m13k5}, \texttt{m17k1} and \texttt{m17k5}, and \texttt{m20k1} and \texttt{m20k5}, they share the same disk mass and similar temperature profiles within each pair, so the $\tau_\mathrm{mid}$ differences reflect the factor-of-five change in $\kappa_0$. However, for the fragmented \texttt{m15k1} disk and the non-fragmented \texttt{m15k5} disk, their differences in temperature profiles and $\kappa_0$ roughly cancel out, resulting in similar $\tau_\mathrm{mid}$, particularly in the outer disk.

The complex interplay between opacity, temperature, and density governs the radial distribution of the dimensionless cooling efficiency $\beta_\mathrm{cool}$ \citep[e.g.][]{Xu_2025ApJ...986...91X} (bottom-left panel). Despite this complexity, a clear dichotomy emerges: non-fragmenting disks maintain $\beta_\mathrm{cool} \gtrsim 10$ across most of the disk, indicating slow cooling consistent with gravito-turbulent regulation. Fragmenting disks, on the other hand, exhibit radial zones where $\beta_\mathrm{cool}$ drops to near unity. We mark the locations of the first clump formation in the five fragmented runs, finding that they lie within regions where $\beta_\mathrm{cool} < 3 $ \citep[e.g.][]{Gammie_2001ApJ...553..174G, Deng_2017ApJ...847...43D,Baehr_2017ApJ...848...40B}.

Finally, the Toomre $Q$ parameter (bottom-right panel) captures the gravitational stability of the disks. In all simulations, $Q$ is initialized to 1.5 at $100$ AU, but subsequent evolution leads to substantial variation. Non-fragmenting disks regulate toward marginal stability, with $Q \sim 1.2-1.5$ over most of the disk. In contrast, fragmenting disks develop pronounced dips in $Q$, falling below the classical threshold of $Q \sim 1$ at radii where clumps begin to emerge.

\subsection{Gravito-turbulence in non-fragmented disks}

\begin{figure*}[ht!]
\plotone{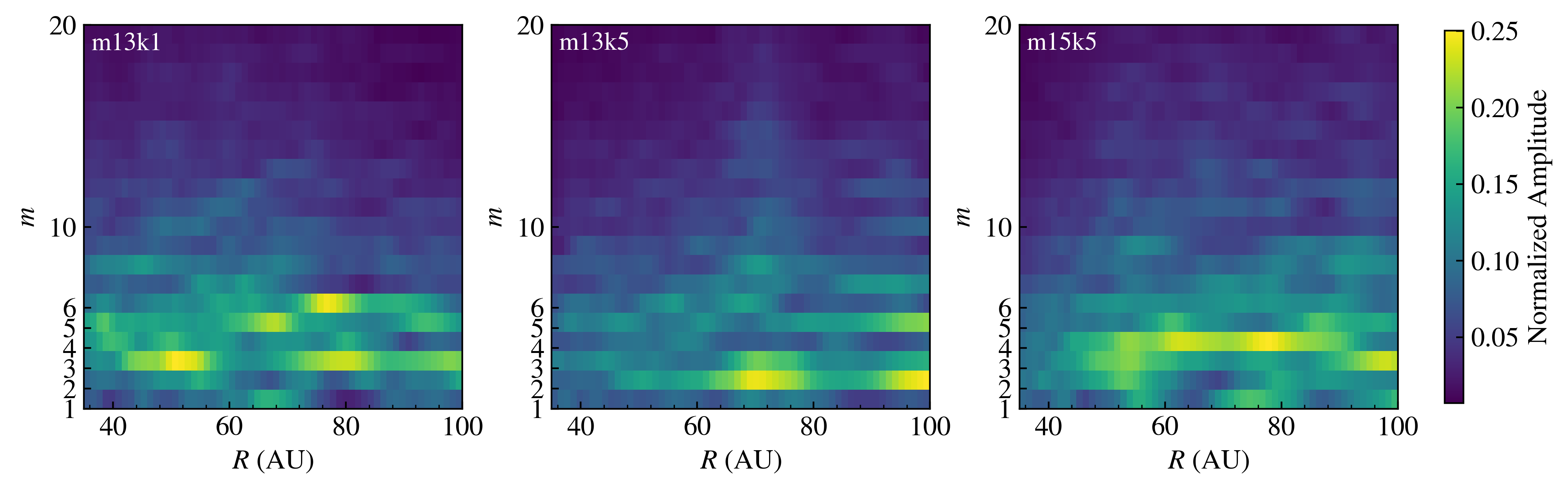}
\caption{Time-averaged Fourier amplitudes of the gas density in the $3$ non-fragmented disks during the $4-5\mathrm{kyr}$. Labels at the upper left corner of each panel indicate the names of these simulation runs. We cover the radius range $35\,\mathrm{AU}<R<100\,\mathrm{AU}$ to exclude the effect of disk outer boundaries. Ticks of $m=1,2,3,4,5,6$ are additionally denoted to show the exact values of the dominated $m$ modes.
\label{fig:phimode_taverage}}
\end{figure*}

We first characterize the gravito-turbulence in the three non-fragmented disks, namely \texttt{m13k1}, \texttt{m13k5}, and \texttt{m15k1}. Figure~\ref{fig:phimode_taverage} displays the time-averaged Fourier amplitudes of the gas density over the period $4$-$5$ $\mathrm{kyr}$ for these disks. In all cases, the density fluctuations are dominated by low-$m$ modes ($m\leq 10$), but the dominant mode varies with radius, reflecting a gradual loss of global coherence in the spiral pattern. For instance, in \texttt{m13k1} the $m=3$ mode is strongest in the inner disk around $55\,\mathrm{AU}$, while at approximately $75\,\mathrm{AU}$ both $m=3$ and $m=6$ modes contribute significantly. In run \texttt{m13k5}, a richer spectrum appears in the outer disk $\sim 100\,\mathrm{AU}$, with notable contributions from $m=2,3,5,7$ modes, indicating the development of more intricate spiral substructures. Similarly, in run \texttt{m15k5} the $m=4$ mode dominates between $60\,\mathrm{AU}$ and $80\,\mathrm{AU}$, with the $m=3$ mode emerging as the prevailing feature beyond $100\,\mathrm{AU}$. Here, the general dominance of low $m$-modes and the radial variations in the Fourier spectra are consistent with other GI simulations \citep[e.g.][]{Cossins_2009MNRAS.393.1157C, Deng_2020ApJ...891..154D}.

\begin{figure*}[ht!]
\plotone{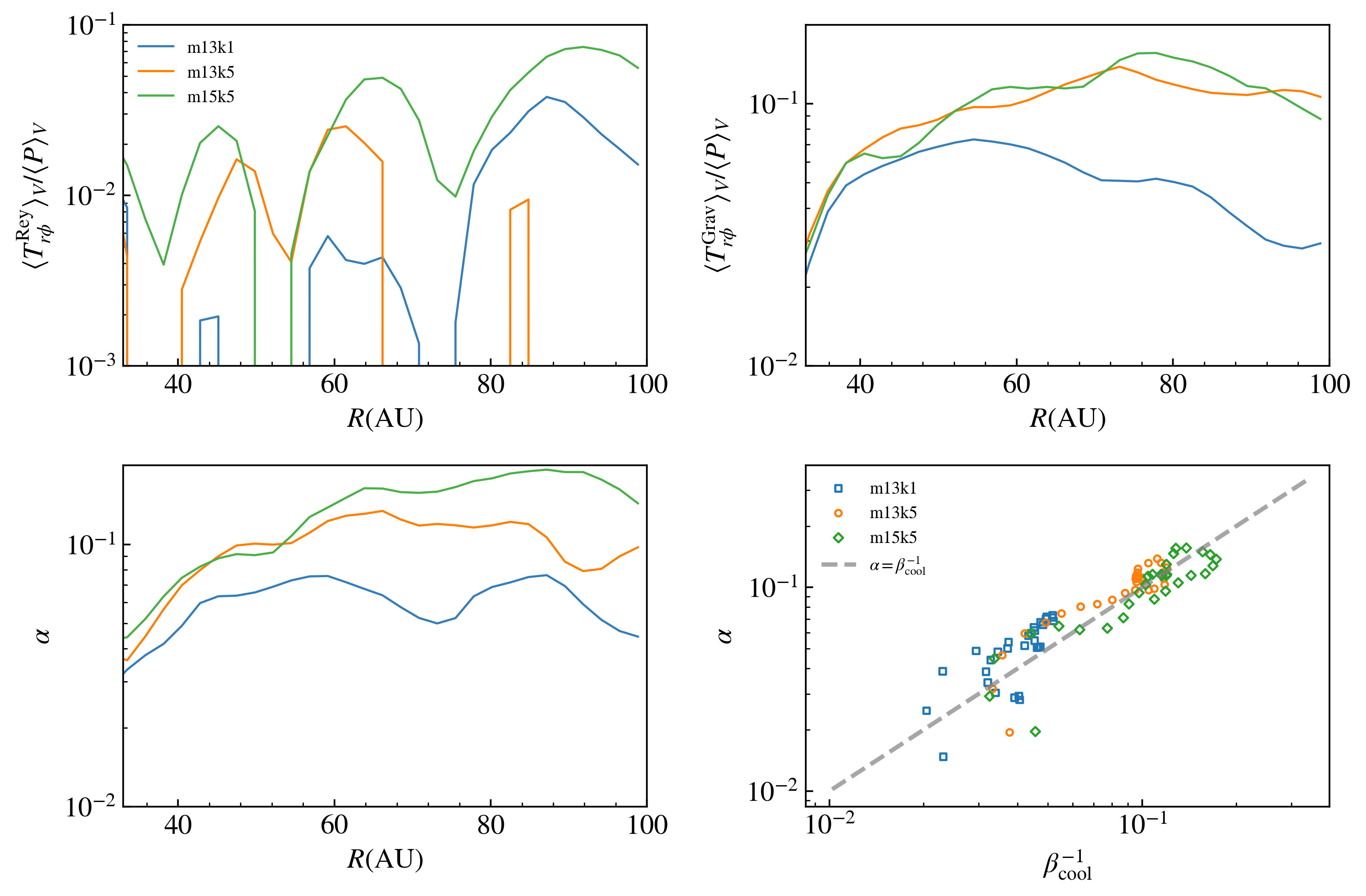}
\caption{Time-averaged Reynolds stress profiles (top left), gravitational stress profiles (top right), viscous $\alpha$ profiles (bottom left) and $\alpha$-$\beta_\mathrm{cool}^{-1}$ relation (bottom right) in the $3$ non-fragmented disks during the $4-5\mathrm{kyr}$. 
\label{fig:stress_beta}}
\end{figure*}

Figure~\ref{fig:stress_beta} compares the time-averaged radial profiles of the Reynolds stress, gravitational stress, and the derived viscous $\alpha$ parameter over the same period. In all non-fragmented runs, the gravitational stress is roughly an order of magnitude larger than the Reynolds stress across all radii, implying that gravito-turbulence drives angular momentum transport primarily via gravitational torques in \texttt{m13k1}, \texttt{m13k5}, and \texttt{m15k1}. The resulting $\alpha$ values are close to $10^{-1}$ in \texttt{m13k5} and \texttt{m15k5}, while \texttt{m13k1} exhibits a slightly lower $\alpha$. We further characterize the relation between $\alpha$ and $\beta_\mathrm{cool}^{-1}$ with $\alpha$ and $\beta_\mathrm{cool}^{-1}$ in all the radial bins in \texttt{m13k1}, \texttt{m13k5}, and \texttt{m15k1}. As discussed in many previous studies \citep[e.g.][]{Gammie_2001ApJ...553..174G, Rice_2005MNRAS.364L..56R, Forgan2017MNRAS.466.3406F, Xu_2025ApJ...986...92X}, the gravito-turbulent steady state follows $\alpha=\frac{3}{2}(\gamma-1)\beta_\mathrm{cool}^{-1} = \beta_\mathrm{cool}^{-1}$. Our simulations confirm this quantitative relation, with the measured $\alpha$ values aligning well with the predicted $\beta_\mathrm{cool}^{-1}$ across the disk.

\subsection{Clump formation and disruption}\label{sec:clumpevolution}

\begin{figure}[ht!]
\plotone{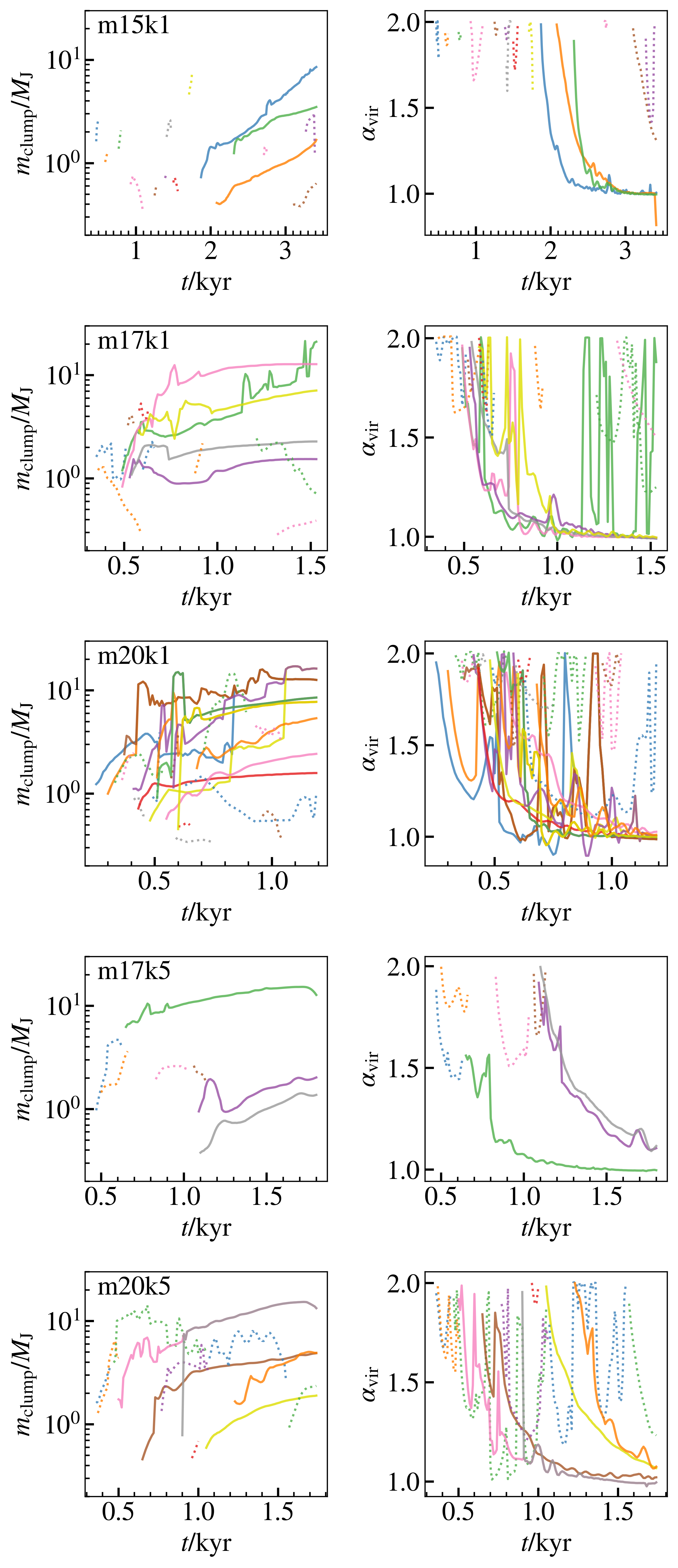}
\caption{Temporal evolution of the clump mass $m_\mathrm{clump}$ (left column) and virial parameter $\alpha_\mathrm{vir}$ (right column) in \texttt{m15k1}, \texttt{m17k1}, \texttt{m20k1}, \texttt{m17k5}, and \texttt{m20k5}. Labels in the upper left of all rows indicate the names of five simulation runs. Each line demonstrates the evolution of an individual clump. Solid lines represent the fragments among all clumps, and dotted lines represent all other clumps. 
\label{fig:fragevolution_massalpha}}
\end{figure}

We now identify and track the evolution of clumps in the fragmented disks, namely \texttt{m15k1}, \texttt{m17k1}, \texttt{m20k1}, \texttt{m17k5}, and \texttt{m20k5}. Figure~\ref{fig:fragevolution_massalpha} shows the temporal evolution of all clumps in these simulations. Generally, more massive disks produce more clumps for a given opacity, while lower-opacity disks produce more clumps at fixed mass, reflecting the more efficient cooling in these environments.

In all fragmented runs, the first generation of clumps appears within a timescale of $\sim1\,\mathrm{kyr}$, roughly one orbit period at $100\,\mathrm{AU}$. Newly formed clumps initially have a virial parameter $\alpha_\mathrm{vir} \sim 2$, indicating a marginally gravitationally bound state. During the first few tens of years of their evolution, $\alpha_\mathrm{vir}$ tends to decrease. However, clumps are subjected to dynamical perturbations from differential shear, collisions with spiral arms, and close encounters with other clumps, all of which can induce fluctuations in $\alpha_\mathrm{vir}(t)$. These perturbations can sometimes be strong enough to disrupt clumps, as indicated by the dotted lines in Figure~\ref{fig:fragevolution_massalpha}. For clumps that eventually dissolve, $\alpha_\mathrm{vir}$ typically shows an initial decline followed by an increase, with the minimum value remaining above unity, and their lifetimes rarely exceed a few hundred years. In contrast, clumps that survive and become bound fragments, shown by the solid lines in Figure~\ref{fig:fragevolution_massalpha}, exhibit a continual decrease in $\alpha_\mathrm{vir}$ toward unity, despite fluctuations caused by interactions with the surrounding gas.

Meanwhile, the evolution of the clump mass $m_\mathrm{clump}$ differs significantly between dissolved clumps and bound fragments. Dissolved clumps exhibit non-monotonic mass evolution due to the complex interplay with disk dynamics. By contrast, bound fragments show a general trend of mass growth, though with occasional fluctuations. The efficiency of mass accretion onto fragments depends sensitively on their migration and the local disk environment, including factors such as gas density, accretion streams, and shear flows, leading to mass evolution that varies widely from gradual, uninterrupted increases to erratic, fluctuation-dominated trajectories as shown in Figure~\ref{fig:fragevolution_massalpha}. A detailed investigation of fragment migration, accretion, and contraction will be presented in a future paper.

\begin{figure*}[ht!]
\plotone{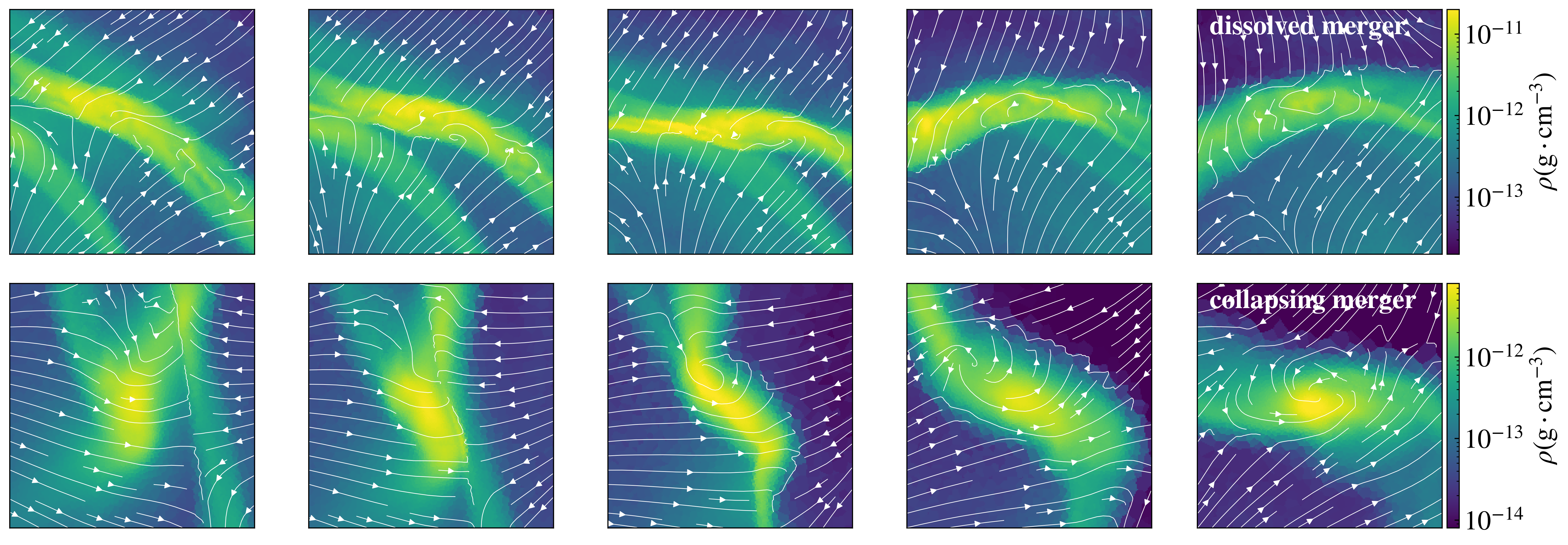}
\caption{Two examples of clumps undergoing disruption or collapse following collisions with spiral arms, taken from simulation \texttt{m17k5}, are shown in the top and bottom panels, respectively. Each row presents a time sequence of mid-plane density slices centered on the clump’s center of mass at different epochs. Streamlines are overplotted in the local Keplerian orbital frame to illustrate the gas motion relative to the background flow. Each slice spans a physical size of $15\,\mathrm{AU} \times 15\,\mathrm{AU}$.
\label{fig:showcase}}
\end{figure*}

Our closer inspection of clump dynamics reveals that dynamical perturbations can either disrupt clumps or promote their collapse, depending on the local balance of forces. In our simulations, we illustrate both possibilities in Figure~\ref{fig:showcase}. The top row shows an example of clump disruption during a clump-spiral arm collision: in the first three snapshots, a dense filamentary structure forms at the edge of the interaction, accompanied by bent streamlines and vortex generation. However, no subsequent contraction occurs, and the clump ultimately dissolves. In contrast, the bottom row presents a case where a clump collides with a spiral arm, triggering the formation of a dense, irregular structure that, after several tens of years, collapses under self-gravity and becomes a bound fragment. This example of clump disruption during a clump-spiral arm collision is broadly consistent with the findings of \citet{Xu_2025ApJ...986...91X}, although their simulations produced no long-lived clumps, so it remains unclear whether collisions can also trigger collapse in their models. More generally, the outcomes of clump-clump, clump-spiral arm, and spiral arm-spiral arm interactions are highly nontrivial, reflecting the delicate competition between thermal pressure, shear, and self-gravity \citep[e.g.,][]{Matzkevich_2024A&A...691A.184M}.

\subsection{Initial fragment mass}

\begin{figure*}[ht!]
\plotone{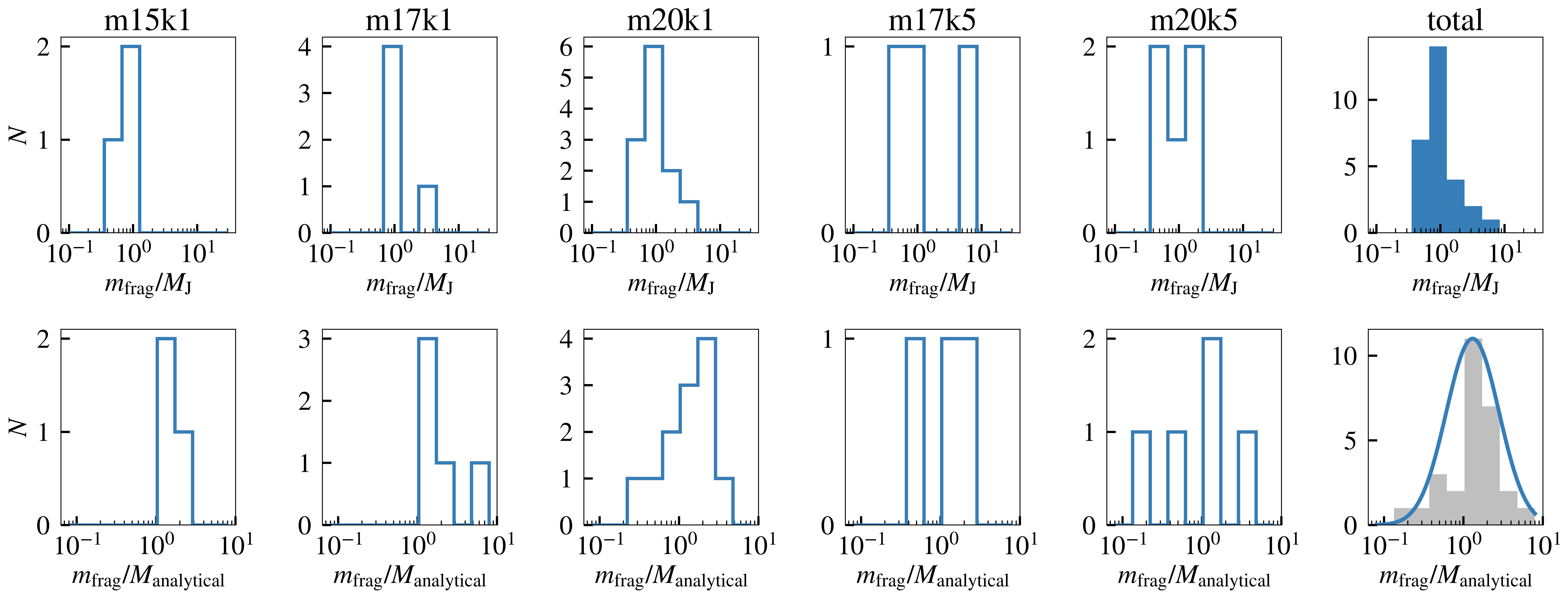}
\caption{The histograms of the initial fragment mass $m_\mathrm{frag}$ in Jupiter mass (top) and the analytically estimated fragment mass $M_\mathrm{analytical}$ (bottom). From left to right, the first five columns are the distribution of the initial fragment mass in \texttt{m15k1}, \texttt{m17k1}, \texttt{m20k1}, \texttt{m17k5}, and \texttt{m20k5}. Labels over the first column indicate the simulations in the five rows. The histogram in the last column stacks the initial fragment mass from all simulations. The solid line in the last column is a log-normal fitting of the distribution with $\sigma [\log(m_\mathrm{frag}/M_\mathrm{analytical})]\sim0.3$.
\label{fig:fragmassdistribution}}
\end{figure*}

In Figure~\ref{fig:fragmassdistribution}, we present the distribution of the initial fragment masses for the five fragmented disks. In physical units, the initial masses span roughly $0.3$--$10\,M_\mathrm{J}$, with a peak around $1\,M_\mathrm{J}$. Although fragments continue to accrete and grow after their formation, as shown in Figure~\ref{fig:fragevolution_massalpha}, the relatively low initial masses suggest that these fragments can potentially evolve into gas giants \citep[e.g.,][]{Boss_1997Sci...276.1836B, Boley_2010Icar..207..509B}. The characteristic masses of order $\sim1\,M_\mathrm{J}$ obtained here are consistent with the RHD simulations employing FLD radiation transport by \citet{Boss_2021ApJ...923...93B}, which found similar fragment masses in disks extending to $20\,\mathrm{AU}$, despite the smaller disk sizes compared to our models ($\gtrsim 100 \,\mathrm{AU}$). While the long-term evolution of these fragments is beyond the scope of this paper, we focus here on understanding the physical origin of the observed Jupiter-mass-scale initial fragment masses.

To theoretically estimate the typical mass scale associated with gravitational fragmentation, a simple approach assumes that the most unstable wavelength $\lambda_\mathrm{T}$ sets the size of the collapsing region. For axisymmetric instabilities, this yields a characteristic mass 
\begin{equation}
M_\mathrm{Toomre} = \pi \left( \frac{\lambda_\mathrm{T}}{2} \right)^2 \Sigma,
\end{equation}
where $\lambda_\mathrm{T} = 2c_s^2/(G\Sigma)$ is the most unstable wavelength \citep[e.g.,][]{Nelson_2006MNRAS.373.1039N}. However, the assumption of axisymmetry breaks down in fragmented disks, where collapse typically occurs within spiral arms, as can be seen in Figure~\ref{fig:RhoProj_multirun}.

To account for this, we adopt an alternative analytic estimate motivated by \citet{Boley_2010Icar..207..509B}, based on collapse within spiral arms:
\begin{equation}\label{eq:fragmass}
    M_\mathrm{analytical} = \Sigma\cdot  \lambda_\mathrm{T} \cdot 2\frac{ c_s}{\Omega_\mathrm{K}},
\end{equation}
The quantities $\Sigma$ and $c_s$ are azimuthally averaged within a radial slice $(R_\mathrm{frag}-R_\mathrm{eff}, R_\mathrm{frag}+R_\mathrm{eff})$, and $\Omega_\mathrm{K}$ is evaluated at $R_\mathrm{frag}$, the heliocentric distance of the fragment at the first moment we identify the fragment.

In the bottom panel of Figure~\ref{fig:fragmassdistribution}, we plot the distribution of initial fragment masses normalized by $M_\mathrm{analytical}$. The resulting distribution, stacked from all fragmented disks, exhibits a clear log-normal shape, peaking around $\log(m_\mathrm{frag}/M_\mathrm{analytical}) \sim \log(1.3)$, with $\sigma [\log(m_\mathrm{frag}/M_\mathrm{analytical})]\sim0.3$. Despite the nonlinear nature of the fragmentation process, the simple $M_\mathrm{analytical}$ estimate based on local disk properties provides a good description for different simulations, suggesting a universal fragmentation scale.

\section{Discussion}\label{sec:discussion}

\subsection{Comparisons with previous studies}

A recent study of the self-gravitating disk using RHD by \citet{Xu_2025ApJ...986...91X} also found that clump masses, although none of them collapsed, follow a log-normal distribution. However, their distribution is centered around $\sim 20\,M_\mathrm{J}$, approximately an order of magnitude larger than the characteristic mass found in our simulations.  This discrepancy is less likely due to treatment of radiation as their RHD simulations and simulations with constant $\beta_\mathrm{cool}$ predict similar mass distributions, which is also the case in our current RHD simulations and previous constant $\beta_\mathrm{cool}$ simulations \citep{Kubli_2023MNRAS.525.2731K,Deng_2021NatAs...5..440D}. The origin of the discrepancy is likely numerical, as \citet{Xu_2025ApJ...986...91X} used the finite volume code with fixed grid and resolution, as opposed to our Lagrangian code with adaptive resolution.

The mass of clumps is known to be sensitive to numerical resolutions as clumps form within thin spirals confined to the mid-plane by self-gravity \citep{Boley_2010Icar..207..509B}. Poor numerical resolution biases toward the formation of large clumps, as light clumps are unable to become gravitationally bound due to limited force resolution. They are also preferably identified in the gravito-turbulent flow because light fragments are prone to disruption either physically or numerically, making them difficult to detect. As a result, the resolution difference likely affected the scale of fragmentation resolved in the simulations. In this work, our quasi-Lagrangian method achieves a resolution sufficient to resolve a $1\,M_\mathrm{J}$ clump with more than $10^{4}$ cells.

Resolving the fragmentation scale does not guarantee resolving the clump evolution, e.g., disruption and collapse (see Figure \ref{fig:showcase}). As noted in \cite{Xu_2025ApJ...986...91X}, around poorly resolved clumps, numerical heating can be strong enough to affect the clump evolution \citep[see also][]{Fromang2005effect}. This highlights the necessity of resolution refinement around clumps. In fact, even with very fast cooling with $\beta_\mathrm{cool}=2$, no clumps collapse and survive as long-lived fragments in \cite{Xu_2025ApJ...986...91X}. However, in similar finite volume simulations with $\beta_\mathrm{cool}<6$, collapsed fragments are observed thanks to adaptive mesh refinement \citep{Brucy2021two}. On the other hand, smoothed particle hydrodynamics simulations as an alternative Lagrangian method form long-lived fragments at $\beta_\mathrm{cool}=4$, similar in mass to our RHD and $\beta$-cooling simulations \citep{Forgan2017MNRAS.466.3406F}.

We also note a slight difference in the definition of the initial fragment mass between the two studies. In \citet{Xu_2025ApJ...986...91X}, the initial mass is measured once the fragment becomes gravitationally bound, whereas in our work, we trace fragments back to the marginally bound stage when collapse begins, and adopt the clump mass at that earlier point as the initial fragment mass. This difference naturally leads to somewhat smaller initial fragment masses in our analysis. However, as shown in Figure~\ref{fig:fragevolution_massalpha}, the mass growth during the transition from marginally bound to gravitationally bound cannot explain the order-of-magnitude difference in the initial fragment mass between the two studies. While we emphasize that our definition more faithfully captures the physical onset of non-linear gravitational collapse, we also confirm that, even under the definition adopted by \citet{Xu_2025ApJ...986...91X}, the initial fragment mass remains well below $10^1\,M_\mathrm{J}$.

\subsection{Neglect of Stellar Irradiation}

Although irradiation from the central protostar can, in principle, affect the thermal balance of young disks, we do not include it in our radiation hydrodynamics simulations. The main reason is that this work represents our first set of well-resolved RHD simulations of gravitationally unstable disks, and a further exploration of irradiation effects is deferred to future studies. Several additional considerations also motivate this choice. First, as shown in Section~\ref{sec:overview}, our simulated disks settle into a quasi-steady, self-regulated gravito-turbulent state with $Q \sim 1$, and the surface density and temperature profiles follow the scalings, which implies a constant aspect ratio rather than a flared geometry. As a result, heating and dynamical thickening in the inner disk can obscure the outer regions from stellar light, reducing the importance of irradiation in the zones where fragmentation is most likely to occur \citep[e.g.,][]{Xu_2021MNRAS.508.2142X, Xu_2023ApJ...954..190X}. Second, the omission of stellar irradiation allows a direct comparison with \citet{Xu_2025ApJ...986...91X}, which adopts the same treatment. Finally, neglecting irradiation also facilitates comparison with idealized models that assume a constant $\beta_\mathrm{cool}$ prescription. If irradiation were included, it would impose a background heating floor that makes it more difficult to isolate the intrinsic role of local cooling in governing fragmentation.

\subsection{Limitations and future perspectives}\label{sec:caveat}

We acknowledge several limitations of this study and outline potential directions for improvement. First, our results should be examined under a more rigorous treatment of radiation transport. In this work, we adopt a grey approximation with a simple power-law opacity. A more realistic approach would include: (i) multi-frequency radiative transfer \citep[e.g.,][]{Jiang_2014ApJS..213....7J, Jiang_2022ApJS..263....4J, He_2024MNRAS.535.3059H}, and (ii) opacity models that capture sublimation effects -- our current power-law opacity likely overestimates the opacity in clump cores once they heat above $\gtrsim 1500,\mathrm{K}$. 
Second, we do not include magneto-hydrodynamic (MHD) effects. PPDs are magnetized and subject to non-ideal MHD processes, which can significantly alter their dynamics. In particular, the gravito-turbulent state can drive a GI dynamo that amplifies the magnetic field \citep[e.g.,][]{Riols_2018MNRAS.474.2212R, Deng_2020ApJ...891..154D, Bethune_2021A&A...650A..49B}, while strong fields can reduce gas accretion onto clumps and thereby modify the initial fragment mass \citep[e.g.,][]{Deng_2021NatAs...5..440D, Kubli_2023MNRAS.525.2731K}. Incorporating non-ideal MHD into future simulations will be essential for assessing these effects.

\section{Conclusions}\label{sec:conclusions}

We have presented a suite of three-dimensional, global RHD simulations of self-gravitating PPDs, using the MFM method with M1 radiation transport. Our simulations explore how disk mass and opacity regulate the gravito-turbulence, the onset of fragmentation, and the properties of bound clumps. Our main conclusions are summarized here:

\begin{itemize}

\item We find that disk fragmentation from GI is controlled by both disk mass and opacity. More massive disks with lower opacity tend to fragment more readily due to more efficient cooling. The boundary between fragmenting and non-fragmenting cases in our simulations is consistent with the critical cooling thresholds of $\beta_\mathrm{cool} \lesssim 3$ reported in previous studies. Disks that remain in a gravito-turbulent state exhibit spiral structures with dominant low-$m$ modes and effective angular momentum transport with $\alpha \sim \beta_\mathrm{cool}^{-1}$, primarily driven by gravitational torques.

\item In fragmented runs, we track the evolution of clumps and identify a population of bound fragments that survive for hundreds of years and reach a virial parameter $\alpha_\mathrm{vir} < 1.1$. These fragments typically form within $\sim 0.5\,\mathrm{kyr}$ and initially exhibit $\alpha_\mathrm{vir} \sim 2$, followed by either collapse or disruption, depending on their interactions with spiral arms and neighboring clumps.

\item We compute the initial fragment mass distribution and find a peak around $1\,M_\mathrm{J}$, well within the planetary mass regime. When normalized by an analytical mass estimate based on local disk conditions, the distribution takes a log-normal form centered at $m_\mathrm{frag}/(\Sigma\lambda_\mathrm{T} \frac{2c_s}{\Omega_\mathrm{K} }) \sim 1.3$ with a scatter of $\sim 0.3$ dex. This points to a universal fragmentation scale tied to local properties, rather than global disk configuration.

\end{itemize}

Our results support the view that gravitational instability in radiatively cooled disks can produce planetary-mass fragments. These findings motivate future work on the long-term evolution of fragments, their migration and accretion, and the role of magnetic fields and irradiation in modifying the fragmentation process.

\begin{acknowledgments}

This work is supported by the National Science Foundation of China under grants No. 12233004, 12325304, 12342501.
H.D. acknowledges support from the Chinese Academy of Sciences via a talent program. YN thanks Lucio Mayer, Yi-Xian Chen, Hai-Yang Wang, and Tianhao Li for valuable discussions. Numerical simulations were performed on Shanhe at the National Supercomputer Center in Jinan.

\end{acknowledgments}

%



\appendix

\section{Benchmark on $M$1-RHD module}\label{sec:app_a}

In this section, we run a series of tests to show the performance of our adapted M$1$-RHD module in GIZMO, mainly the iterative implicit solver for stiff radiation source terms, and the HLL Riemann solver with piecewise linear reconstruction for the radiation transport.

\subsection{Radiation-Matter coupling}\label{sec:coupling}

We first verify the correctness of our implicit iterative stiff source solver on an essentially zero-dimensional problem, a standard test of the radiation subsystem \citep[e.g.][]{Turner_2001ApJS..135...95T, MelonFuksman_2019ApJS..242...20M}. In this test, we consider a static uniform absorbing fluid initially out of thermal balance, and set up the radiation field so that radiation energy dominates the total energy. In this way, the radiation energy emitted or absorbed in reaching equilibrium is a small fraction of the initial value, so radiation energy density $E_\mathrm{r}$ is assumed constant. As a result, an analytical solution can be obtained in this test.

\begin{figure}[ht!]
\centering
\includegraphics[width=0.5\columnwidth]{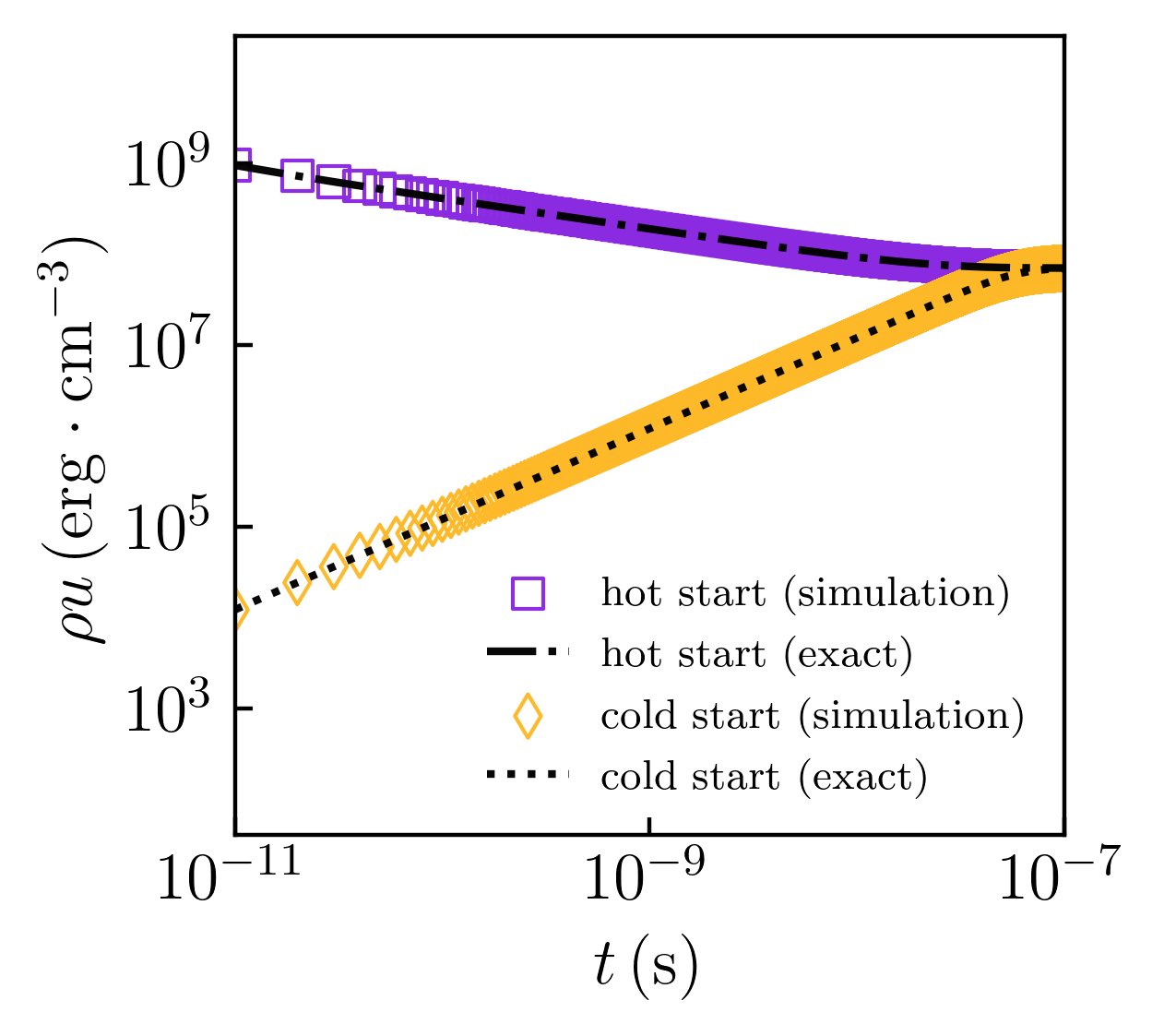}
\caption{Gas internal energy density as a function of time in two simulations and the corresponding analytical solution.}
\label{fig:test_coupling}
\end{figure}

We run the test with two initial conditions: (a)``hot start'' case, where the initial gas internal energy density $\rho u=10^{10}\,\mathrm{erg}\cdot{cm}^{-3}$; (b)``cold start'' case, where the initial gas energy density $\rho u=10^{2}\,\mathrm{erg}\cdot{cm}^{-3}$. In both two cases, we adopt a gas density $\rho=10^{-7}\,\mathrm{g}\cdot{cm}^{-3}$, a radiation energy density $E_\mathrm{r}=10^{12} \,\mathrm{erg}\cdot{cm}^{-3}$, an opacity $\kappa=0.4 \,\mathrm{cm}\cdot \mathrm{g}^{-1}$, a mean molecular weight $\mu=0.6$, and an equation of state gamma $\gamma=5/3$. We note that although this test problem is essentially zero-dimensional, we use four cells in one dimension to run the test, which is the minimal requirement for an appropriate kernel establishment in GIZMO. We run both of the simulation till $t=10^{-7}\,\mathrm{s}$, and the output snapshots are saved at a cadence of $\Delta t=10^{-11}\,\mathrm{s}$. As shown in Figure~\ref{fig:test_coupling}, the numerical solutions match the analytical ones well in the considered time range, i.e. $10^{-11}\,\mathrm{s}<t<10^{-7}\,\mathrm{s}$.

\subsection{Radiation wave propagation}\label{sec:wave2d}

We now verify our HLL Riemann solver with piecewise linear reconstruction using the propagation of small-amplitude, free-streaming radiation waves in a purely absorbing, homogeneous medium with low optical depth, which has become a standard test in the hyperbolic transport problems \citep[e.g.][]{Gardiner_2005JCoPh.205..509G, Skinner_2013ApJS..206...21S, Kannan_2019MNRAS.485..117K, Fuksman_2021ApJ...906...78M}.

We initialize a two-dimensional periodic box of size $(L_x,L_y)=(2,1)$. The initial radiation field is given by 
\begin{equation}
    E_\mathrm{r}(\bm{r},t=0) = E_\mathrm{bkg}+\epsilon \sin(\frac{2 \pi \bm{k}\cdot \bm{r}}{\lambda}),
\end{equation}
\begin{equation}
    \bm{F}_\mathrm{r}(\bm{r},t=0) = cE_\mathrm{r}(\bm{r},t=0)\bm{k},
\end{equation}
where $E_\mathrm{bkg} = 1$ is the constant and uniform energy density background, $\epsilon=10^{-6}$ is the initial wave amplitude, $\lambda=\frac{2}{\sqrt{5}}$ is the wave length, $\bm{k}=(1/\sqrt{5},2/\sqrt{5})$ is the wave vector, and $\bm{r}=(x,y)$ is the position vector. We set the opacity and gas density so that the optical depth per wavelength is $\rho \kappa \lambda =0.1$.

We perform a suite of simulations of resolution $2\mathcal{N}\times\mathcal{N}$, with $\mathcal{N}$ covering $8,16,32,64,128$. Each simulation is run till $t_\lambda=\frac{\lambda}{c}$, when the wave propagates a distance of one wavelength. We compare the radiation energy density to the analytic solution 
\begin{equation}
    E^\mathrm{*}_\mathrm{r}(\bm{r},t=t_\lambda) = E_\mathrm{bkg}+\epsilon e^{-1} \sin(\frac{2 \pi \bm{k}\cdot \bm{r}}{\lambda}),
\end{equation}
and calculate the $L_1$-error defined as
\begin{equation}
    |\delta E_\mathrm{r}| = \frac{1}{2\mathcal{N}^2}\sum_i |E_{\mathrm{r}}(\bm{r}_i,t=t_\lambda)-E^\mathrm{*}_{\mathrm{r}}(\bm{r}_i,t=t_\lambda)|,
\end{equation} where the summation runs over the whole box.

\begin{figure}[ht!]
\centering
\includegraphics[width=0.5\columnwidth]{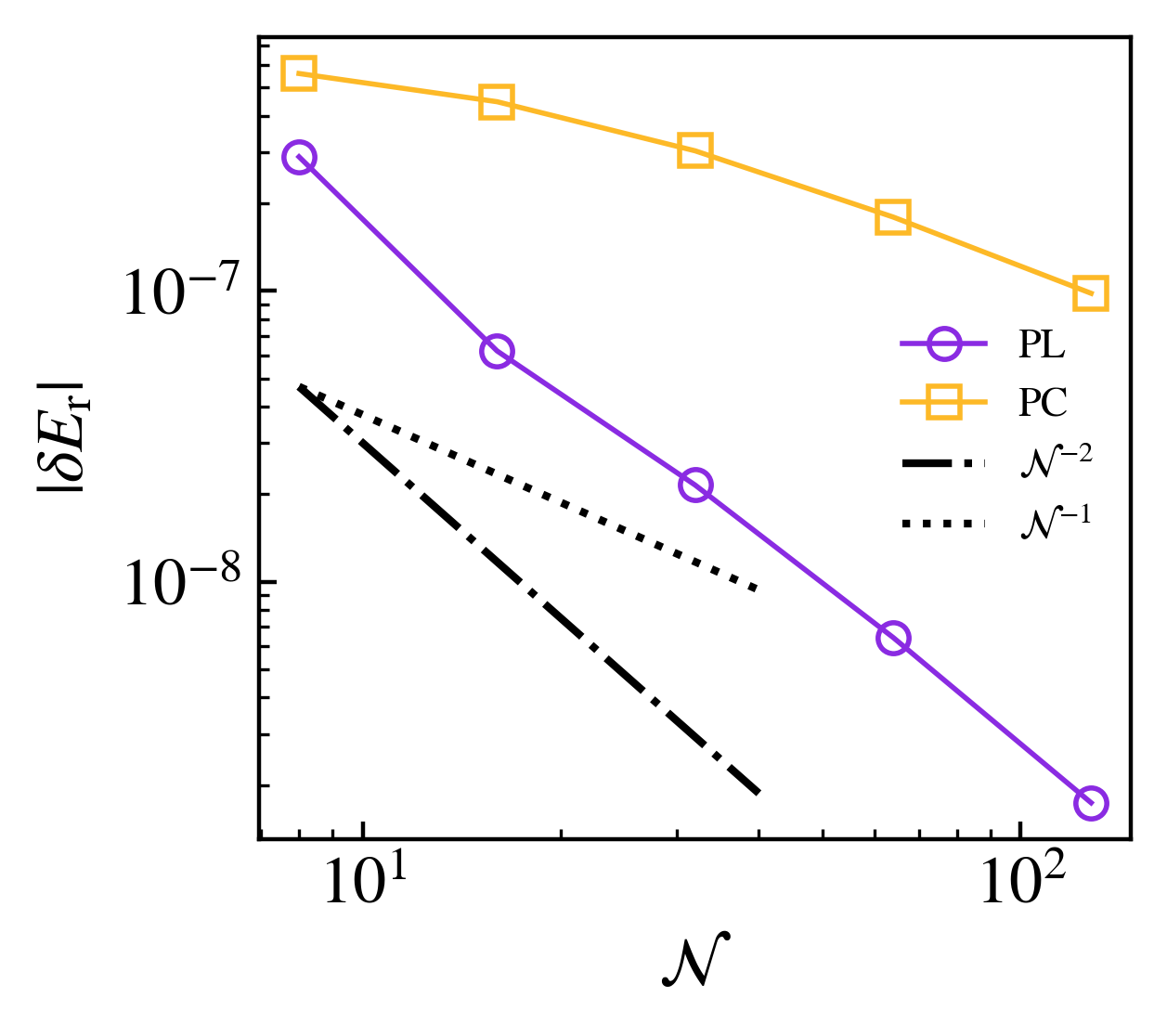}
\caption{Convergence of $|\delta E_\mathrm{r}|$ yielding various resolutions of the radiation wave propagation test. The yellow line represents tests with piecewise linear reconstruction, and the purple line represents tests with piecewise constant reconstruction. Additionally, a dotted black line indicating $\mathcal{N}^{-1}$ scaling and a dash-dotted black line indicating $\mathcal{N}^{-2}$ scaling are overplotted for reference.
\label{fig:test_wavescaling}}
\end{figure}

To better show the higher accuracy achieved by our piecewise linear reconstruction in the HLL Riemann solver, we also run the same test without reconstruction, i.e., piecewise constant reconstruction. We label the former scheme as ``PL'' and the latter as ``PC''. Figure~\ref{fig:test_wavescaling} shows nearly second-order convergence rate of our HLL Riemann solver with piecewise linear reconstruction, which is one order higher than the HLL Riemann solver without reconstruction.

\subsection{Shadows}\label{sec:shadow}

To further assess the performance of our radiation hydrodynamics solver, we consider a classical test problem designed to evaluate the ability of moment-based radiation transport schemes to handle anisotropic radiation fields and cast shadows. This test is especially suited to highlight the advantages of the M1 closure over lower-order approximations like flux-limited diffusion (FLD), which tend to artificially smooth angular features of the radiation field. Specifically, we adopt a commonly used two-dimensional setup in which a collimated radiation beam impinges upon a dense, opaque cloud embedded in a diffuse medium. Similar shadowing tests have been carried out by several authors to benchmark M1 methods in other hydrodynamical codes \citep[e.g.][]{Gonzalez_2007A&A...464..429G, Skinner_2013ApJS..206...21S, MelonFuksman_2019ApJS..242...20M}.

We follow a widely used setup in which a dense, elliptical cloud is embedded in a lower-density ambient medium within a two-dimensional Cartesian domain of size $ (L_x, L_y) = (1.0\,\mathrm{cm}, 0.24\,\mathrm{cm})$. The ambient gas has a uniform density $ \rho_0 = 10^{-3} \,\mathrm{g\cdot cm^{-3}} $, while the cloud has a central density $ \rho_1 = 1.0 \,\mathrm{g \cdot cm^{-3}} $ and a smooth density distribution given by
\begin{equation}
\rho(x,y) = \rho_0 + \frac{\rho_1 - \rho_0}{1 + e^\Delta},
\end{equation}
where
\begin{equation}
\Delta = 10\left[\left(\frac{x - x_c}{x_0}\right)^2 + \left(\frac{y - y_c}{y_0}\right)^2 - 1\right].
\end{equation}
This describes a cloud with smooth edges centered at $ (x_c, y_c) = (0.5, 0.06) $, and semi-axes $ x_0 = 0.10 \,\mathrm{cm} $, $ y_0 = 0.12 \,\mathrm{cm} $. Both the gas and radiation fields are initially in thermal equilibrium at a uniform temperature $ T_0 = 290\,\mathrm{K} $.

At $ t = 0 $, a collimated radiation flux corresponding to a blackbody temperature $ T_1 = 6T_0 = 1740\,\mathrm{K} $ is injected uniformly from the left boundary and propagates along the $ +x $ direction. The opacity is taken to be temperature- and density-dependent:
\begin{equation}
\kappa(\rho, T) = \kappa_0 \left(\frac{\rho}{\rho_0}\right) \left(\frac{T}{T_0}\right)^{-3.5},
\end{equation}
where $ \kappa_0 = 100\,\mathrm{cm^2\cdot \,g^{-1}} $. This ensures the ambient medium is optically thin while the dense cloud remains optically thick to incoming radiation. The bottom and top boundaries are set to be periodic, and the right boundary allows outflow. We initialize the domain using a resolution of $280 \times 160$ cells and evolve the system for $10$ light-crossing times across the horizontal extent of the box. To demonstrate the improvements brought by piecewise linear reconstruction in our HLL Riemann solver, we perform a set of runs with different solver configurations. Specifically, we vary both the numerical flux function (HLL vs. GLF) and the reconstruction scheme (piecewise constant vs. piecewise linear). We denote the methods as follows: ``HLL-PL'' (fiducial) and ``HLL-PC'' for the HLL solver with piecewise linear and constant reconstruction, respectively, and ``GLF-PL'' and ``GLF-PC'' for the GLF solver with linear and constant reconstruction.

\begin{figure}[ht!]
\centering
\includegraphics[width=0.5\columnwidth]{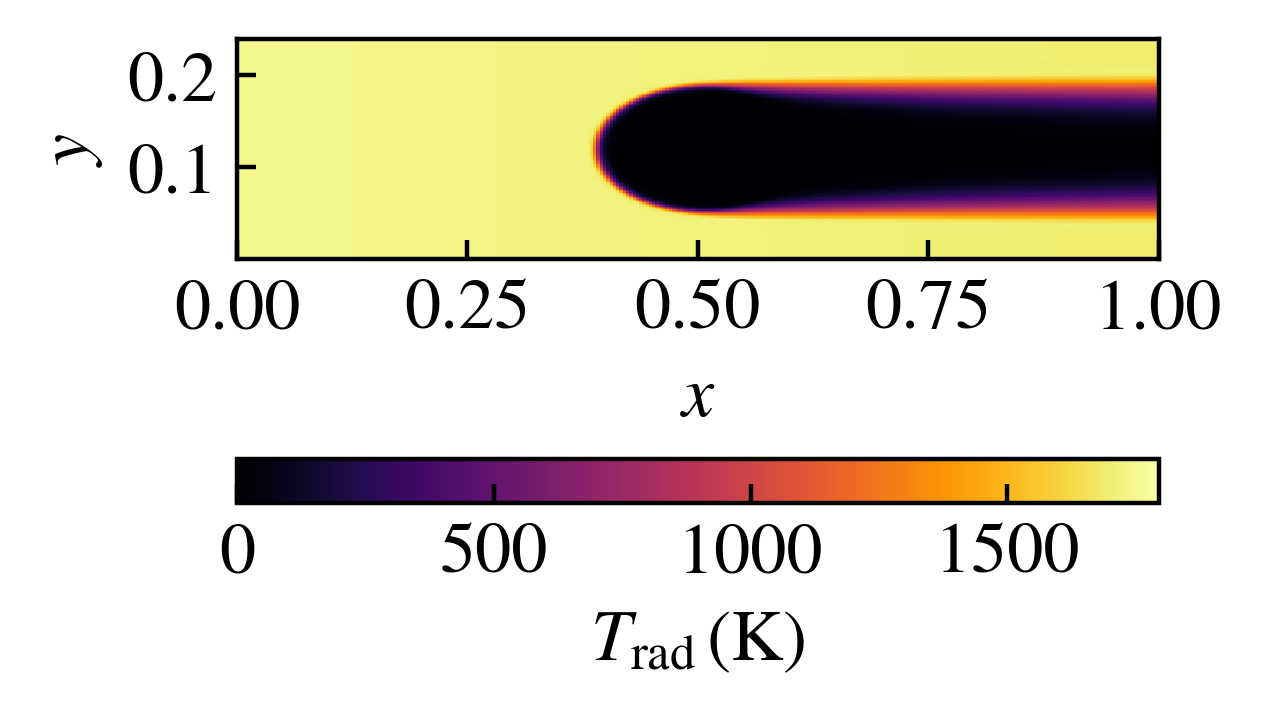}
\caption{Two-dimensional distribution of the radiation temperature $T_\mathrm{rad}=(E_r/a_r)^{1/4}$ in the HLL-PL run after $10$ light-crossing times across the horizontal extent of the box.
\label{fig:test_shadow1}}
\end{figure}

\begin{figure}[ht!]
\centering
\includegraphics[width=0.5\columnwidth]{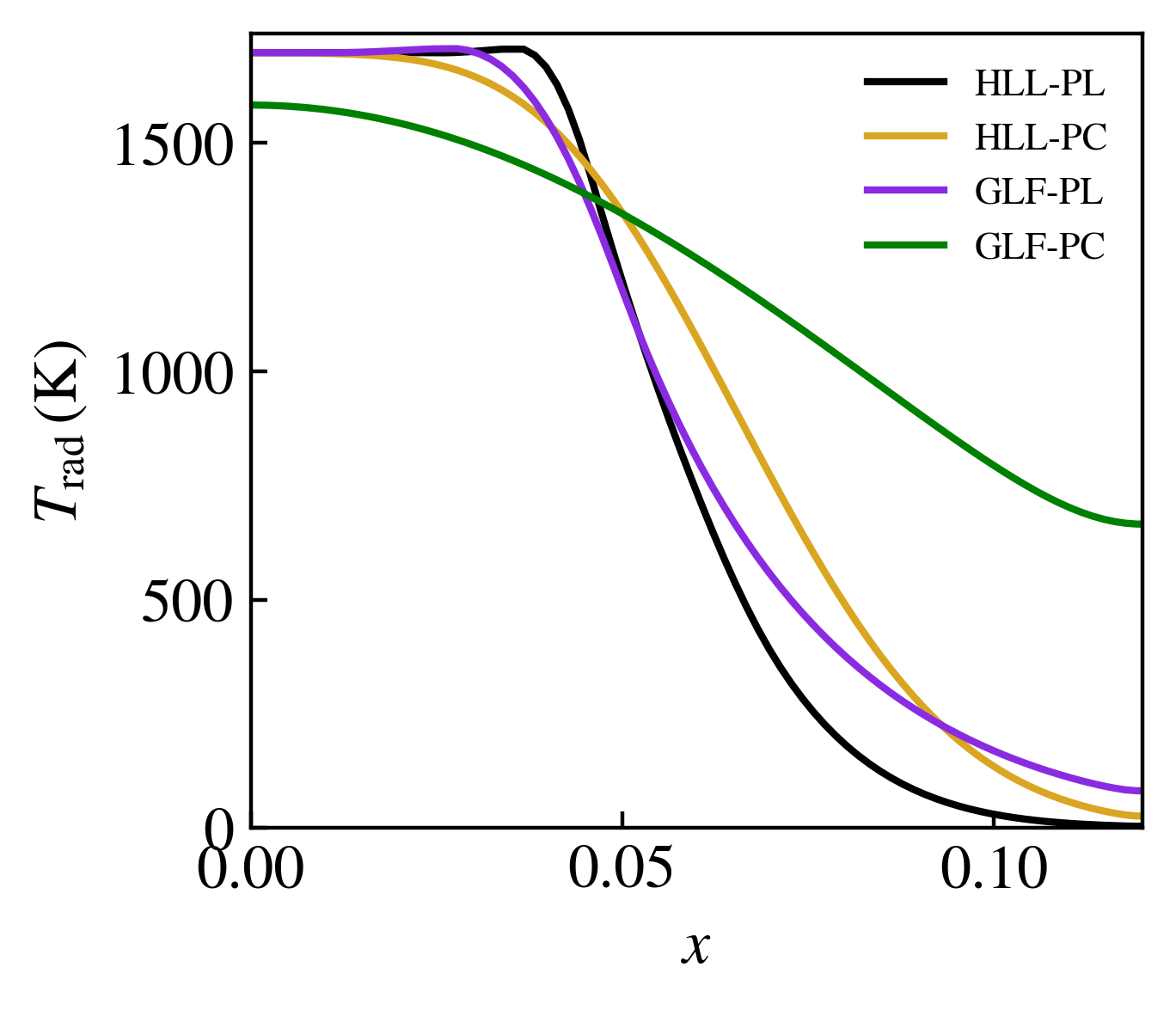}
\caption{Radiation temperature $T_\mathrm{rad}$ measured at the right edge of the box after $10$ light-crossing times across the horizontal extent of the box. Lines of different colors indicate results for different runs, namely HLL-PL (black), HLL-PC (golden), GLF-PL (purple), and GLF-PC (green).
\label{fig:test_shadow2}}
\end{figure}

To evaluate the performance of our M1 radiation module, we examine the resulting radiation fields. As shown in Figure~\ref{fig:test_shadow1}, a well-defined shadow is cast behind the dense cloud, demonstrating the solver’s ability to preserve sharp angular features over a significant distance. In Figure~\ref{fig:test_shadow2}, we compare the shadow sharpness across different solver configurations: HLL-PL produces the sharpest shadow boundary, while GLF-PC shows the most diffusive result, while the HLL-PC and GLF-PL runs yield intermediate levels of diffusion.

\subsection{Radiative Shocks}\label{sec:shadow}

Finally, we couple our radiation subsystem with the hydrodynamical evolution of the matter, and test the code’s ability to reproduce shock waves in optically thick media. 

We adopt the classic one-dimensional setup originally described by \citet{Ensman_1994ApJ...424..275E}, which has since become a standard test in many radiation hydrodynamics studies \citep[e.g.][]{Ensman_1994ApJ...424..275E, Hayes_2003ApJS..147..197H, Gonzalez_2007A&A...464..429G, Fuksman_2021ApJ...906...78M}. The computational domain extends over the interval $x \in [0,\, 7 \times 10^{10}]\,\mathrm{cm}$, and is initially filled with uniform gas of density $\rho_1 = 7.78 \times 10^{-10}\,\mathrm{g \cdot cm^{-3}}$. Both the gas and radiation fields are initialized in local thermodynamic equilibrium (LTE) at a temperature $T_1 = 10\,\mathrm{K}$, with mean molecular weight $\mu = 1$ and adiabatic index $\Gamma = 7/5$. The interaction between radiation and matter is governed by an absorption opacity specified to yield $\kappa \rho = 3.1 \times 10^{-10}\,\mathrm{cm^{-1}}$. To generate a radiative shock, a uniform negative velocity $u$ is imposed throughout the domain, causing material to flow toward the left boundary. A reflecting boundary condition is applied at the left edge to initiate a rightward-propagating shock. The shock velocity is set to $u = -6\,\mathrm{km\cdot s^{-1}}$ for the subcritical case and $u = -20\,\mathrm{km\cdot s^{-1}}$ for the supercritical case. In both cases, we use $1024$ cells in the domain. For the subcritical case, we adopt the reduced speed of light approximation (RSLA) with a ratio of $\tilde{c}/c = 0.001$. For the supercritical case, we use a less aggressive reduction with $\tilde{c}/c = 0.1$. These choices are motivated by the findings of \citet{Fuksman_2021ApJ...906...78M}, who showed that such values introduce negligible deviations from full-speed-of-light solutions for similar radiative shock problems.

\begin{figure*}[ht!]
\centering
\includegraphics[]{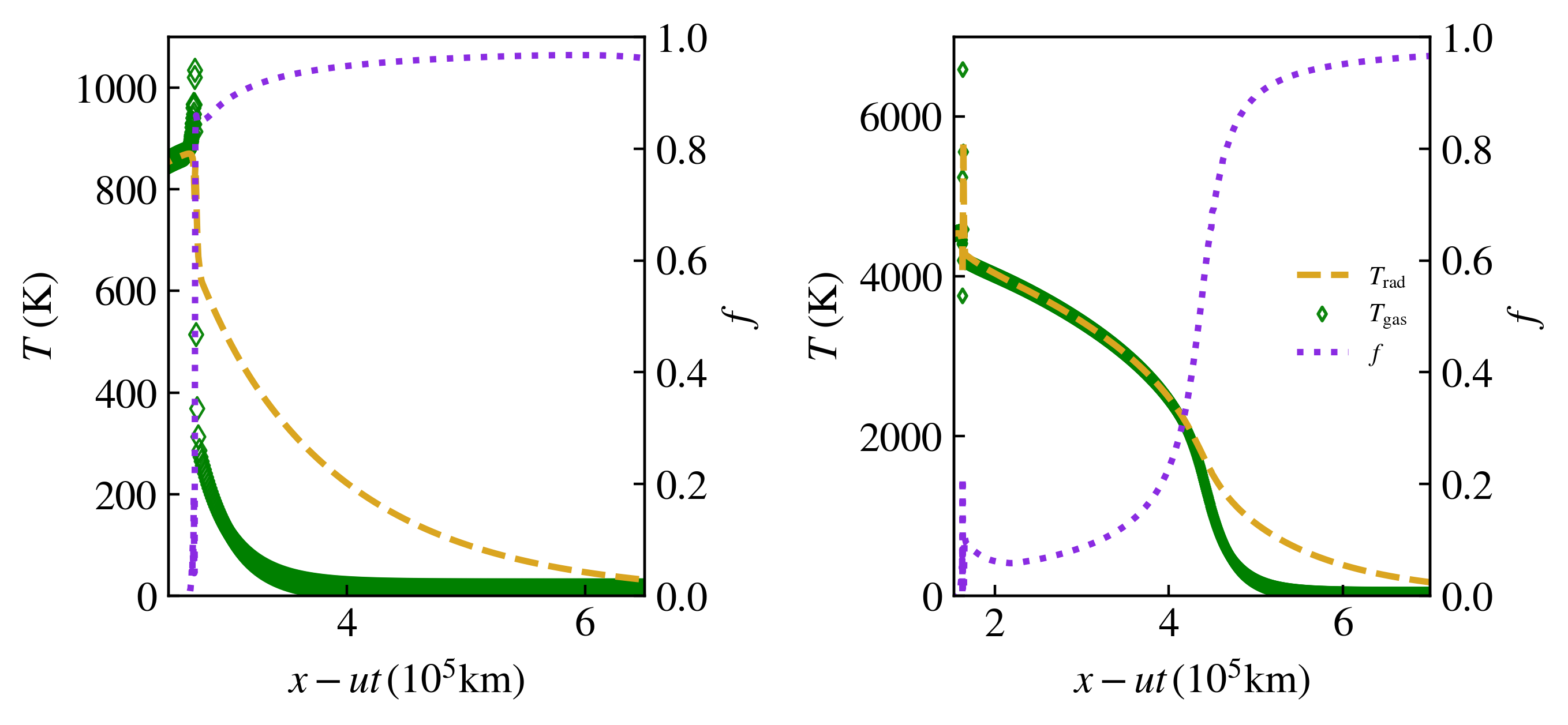}
\caption{Gas and radiation temperature profiles, denoted by $T_{\mathrm{gas}}$ and $T_{\mathrm{rad}}$, for the subcritical (left) and supercritical (right) radiative shock tests. The profiles are shown at times $t = 3.8 \times 10^4\,\mathrm{s}$ and $t = 7.5 \times 10^3\,\mathrm{s}$, respectively, as a function of the comoving coordinate $s = x - ut$. The Reduced radiation flux $f$ is overplotted to illustrate the transition between the free-streaming and diffusion regimes across the shock structure.
\label{fig:test_shock}}
\end{figure*}

In both cases, as illustrated in Figure~\ref{fig:test_shock}, our results are in excellent agreement with the high-resolution ($2048$-cell), no-RSLA reference solutions presented by \citet{Fuksman_2021ApJ...906...78M}. The shock structure, including the radiative precursor, temperature plateau, and post-shock relaxation, is well reproduced. We do, however, observe a small numerical artifact in the form of oscillations in $T_{\mathrm{rad}}$ near the shock peak in the supercritical case. These may result from sharp gradients in radiation energy density combined with the strong coupling to gas temperature in optically thick conditions.

\section{Convergence}\label{sec:app_b}

To directly assess numerical convergence, we re-run the fiducial \texttt{m17k5} simulation starting from the same initial conditions, evolving each run for $1\,\mathrm{kyr}$ (one orbital period at $100\,\mathrm{AU}$). We test two modifications: one with four times the spatial resolution, and the other with four times the reduced speed of light $\tilde{c}$. While the fiducial run evolves for a total of $1.8\,\mathrm{kyr}$, the final $0.8\,\mathrm{kyr}$ becomes increasingly computationally demanding due to strong timestep constraints imposed by collapsing clumps. As a result, extending the high-resolution or high-$\tilde{c}$ runs to the full duration is computationally prohibitive. Our goal is not to capture the long-term evolution in these test runs, but rather to examine how resolution and $\tilde{c}$ influence the early disk evolution and fragmentation. Since clumps begin to form within the first $0.45\,\mathrm{kyr}$ and noticeable differences already emerge between the runs during this phase, we argue that a $1\,\mathrm{kyr}$ integration is sufficient to capture the key convergence behavior.

\begin{figure*}[ht!]
\plotone{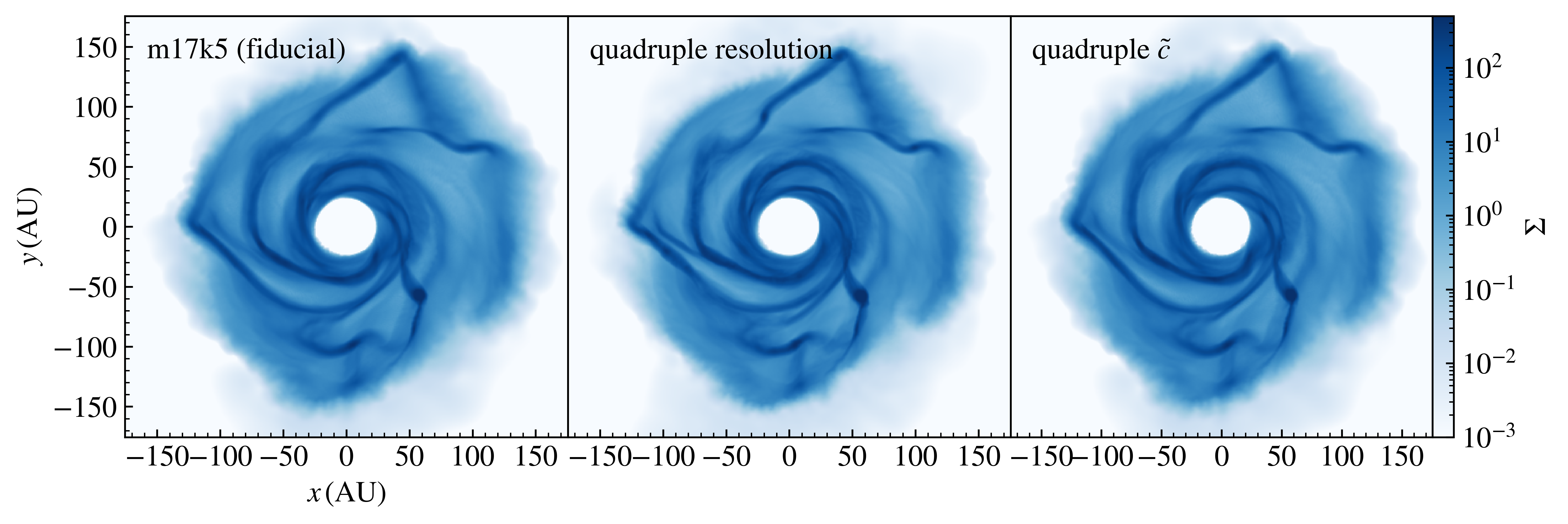}
\caption{Face-on view of the gas surface density projection at $1\,\mathrm{kyr}$ for the fiducial run (left), the quadruple resolution run (middle), and the quadruple $\tilde{c}$ run (bottom).
\label{fig:convergence_proj}}
\end{figure*}

Figure~\ref{fig:convergence_proj} shows the face-on surface density projections of the fiducial run, the quadruple-resolution run, and the quadruple-$\tilde{c}$ run at $t = 1\,\mathrm{kyr}$. All three simulations exhibit similar large-scale spiral structures and global disk morphology. The run with increased $\tilde{c}$ yields results that are nearly indistinguishable from the fiducial case, confirming that the fiducial value of $\tilde{c}$ is sufficient for capturing the relevant radiation dynamics. The high-resolution run, on the other hand, reveals more intricate features within the spiral arms and resolves additional filamentary substructure, particularly in regions of higher density, while the large-scale structures of spiral arms are similar to the fiducial run.

\begin{figure}[ht!]
\centering
\includegraphics[width=0.5\columnwidth]{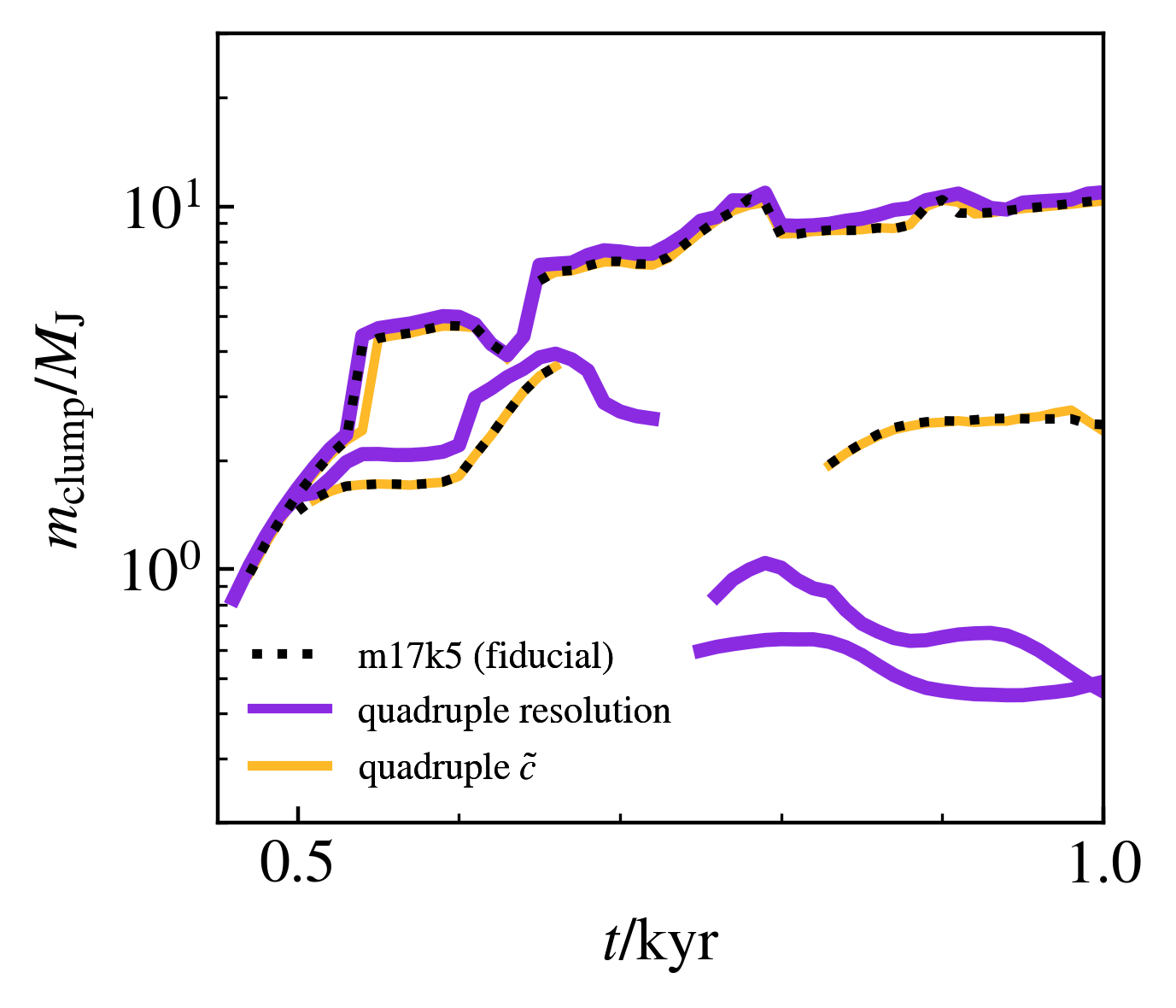}
\caption{Temporal evolution of the clump mass $m_\mathrm{clump}$ within $1\,\mathrm{kyr}$ in the fiducial run (black dotted), the quadruple resolution run (purple solid), and the quadruple $\tilde{c}$ run (yellow solid).
\label{fig:convergence_clump}}
\end{figure}

To quantify the impact on disk fragmentation, we apply our clump tracking algorithm to both convergence runs and compare the resulting clump populations to those of the fiducial simulation. As shown in Figure~\ref{fig:convergence_clump}, the clump evolution in the quadruple-$\tilde{c}$ run closely tracks that of the fiducial run, with nearly identical clump formation times and masses. In contrast, the high-resolution run produces additional low-mass clumps that are not identified in the fiducial simulation, likely due to improved spatial resolution of the fragmentation process.

\bibliography{sample7}{}
\bibliographystyle{aasjournalv7}
\end{CJK*}


\end{document}